\documentclass[12pt,a4paper]{article}
\pdfoutput=1
\usepackage{adjustbox}
\usepackage{color}
\usepackage{amssymb,amsmath,bm,bbm}
\usepackage{epsf}
\usepackage{epsfig}
\usepackage{afterpage}
\usepackage{longtable}
\usepackage[dvipsnames]{xcolor}
\usepackage[linktoc=page,bookmarks=false,colorlinks=false,
linkbordercolor=RoyalBlue,citebordercolor=ForestGreen,urlbordercolor=CornflowerBlue]{hyperref}
\usepackage{latexsym,mathrsfs,dsfont}
\usepackage[normalem]{ulem} % for strikeout \sout
\usepackage[compress]{cite}
\usepackage{graphicx}
\usepackage{url}
\usepackage{paralist}
\usepackage{bbold}

\definecolor{green1}{rgb}{0.06,0.66,0.06}
\definecolor{red}{cmyk}{0,1,1,0.4}

\definecolor{darkgreen}{rgb}{0.0,0.6,0.0}
\definecolor{darkblue}{RGB}{12,13,115}
\definecolor{darkred}{RGB}{204,6,0}

\usepackage{slashed}

\newcommand{\IM}{{\rm Im}}

\newcommand{\mev}{\, {\rm MeV}}

\newcommand{\vcb}{|V_{cb}|}
\newcommand{\vtd}{|V_{td}|}
\newcommand{\vub}{|V_{ub}|}
\newcommand{\vts}{|V_{ts}|}
\newcommand{\vus}{|V_{us}|}

\def\epe{\varepsilon'/\varepsilon}
\newcommand{\beq}{\begin{equation}}
\newcommand{\eeq}{\end{equation}}
\newcommand{\be}{\begin{equation}}
\newcommand{\ee}{\end{equation}}
\newcommand{\bi}{\begin{itemize}}
\newcommand{\ei}{\end{itemize}}
\newcommand{\ba}{\begin{array}}
\newcommand{\ea}{\end{array}}
\newcommand{\beqa}{\begin{eqnarray}}
\newcommand{\eeqa}{\end{eqnarray}}
\newcommand{\bea}{\begin{eqnarray}}
\newcommand{\eea}{\end{eqnarray}}
\newcommand{\beqn}{\begin{eqnarray}}
\newcommand{\eeqn}{\end{eqnarray}}

\newcommand{\eps}{\epsilon}

\definecolor{red}{cmyk}{0,1,1,0.4}

\def\kpn{K^+\rightarrow\pi^+\nu\bar\nu}
\def\klpn{K_{L}\rightarrow\pi^0\nu\bar\nu}

\def\ksm{K_S\to\mu\bar\mu}

\newcommand{\muLow}{{\mu_\text{had}}}
\newcommand{\muEW}{{\mu_\text{ew}}}

\usepackage{tikz}
\usepackage{xcolor}
\definecolor{myblue}{RGB}{21,113,173}
\definecolor{mygreen}{RGB}{16,140,75}

\newcommand{\filledarrowb}{%
\mathrel{%
\begin{tikzpicture}[baseline=-0.5ex, scale=0.5]
\fill[myblue]
(0,0) -- (0.6,0.4) -- (0.6,0.15) -- (2.4,0.15) -- (2.4,0.4) -- (3,0)
-- (2.4,-0.4) -- (2.4,-0.15) -- (0.6,-0.15) -- (0.6,-0.4) -- cycle;
\end{tikzpicture}
}%
}

\newcommand{\filledarrowg}{%
\mathrel{%
\begin{tikzpicture}[baseline=-0.5ex, scale=0.5]
\fill[mygreen]
(0,0) -- (0.6,0.4) -- (0.6,0.15) -- (2.4,0.15) -- (2.4,0.4) -- (3,0)
-- (2.4,-0.4) -- (2.4,-0.15) -- (0.6,-0.15) -- (0.6,-0.4) -- cycle;
\end{tikzpicture}
}%
}

\begin{document}

\begin{flushleft}
{\em Version of \today}
\end{flushleft}

\vspace{-14mm}
\begin{flushright}
        {AJB-26-2}
\end{flushright}

\vspace{8mm}

\begin{center}
{\LARGE\bf
  \boldmath{Hunting New Animalcula with  Flavour Changing Processes}\footnote{
To  appear in the conference proceedings of the 66th Cracow School of Theoretical Physics, in Acta Physica Polonica B, dedicated to Andrzej Białas on the occasion of his 90th birthday.}
\\[8mm]
{\large\bf Andrzej~J.~Buras \\[0.3cm]}}
{\small 
      TUM Institute for Advanced Study, Lichtenbergstr.~2a, D-85748 Garching, Germany\\
      Physik Department, TUM School of Natural Sciences, TU M\"unchen,\\ James-Franck-Stra{\ss}e, D-85748 Garching, Germany\\
      E-mail: aburas@ph.tum.de }
\end{center}

\vspace{4mm}

\begin{abstract}
  \noindent
This year marks the 350th anniversary of the discovery of the first animalcula (little animals)  by van Leeuvanhoek in 1676.
Flavour physics makes it possible to search for new animalcula at distance scales far shorter than those resolved by van Leeuwenhoek in 1676 and even shorter than those directly accessible at the Large Hadron Collider and the planned
colliders in this century. I summarize various strategies for achieving this
goal. While precise measurements of a wide variety of observables 
and their precise theoretical calculations, both within the Standard Model (SM) and beyond it, are indispensable in this context, 
in my view it is crucial to develop strategies for the search for New Physics (NP) that go beyond the global fits that are very popular today. While effective field theories such as WET and SMEFT are formulated in terms of Wilson coefficients of the relevant operators, with correlations characteristic of the SM and of specific NP scenarios, the most direct tests of the SM and its extensions are, in my opinion, correlations among different observables, like branching ratios of numerous decays, that are characteristic of particular new animalcula at work. 
\end{abstract}

\vspace{4mm}
\setcounter{page}{0}
\thispagestyle{empty}
\newpage

\tableofcontents

\section{Overture and Outline}
\begin{figure}[t]
\centering%
\includegraphics[width=0.8\textwidth]{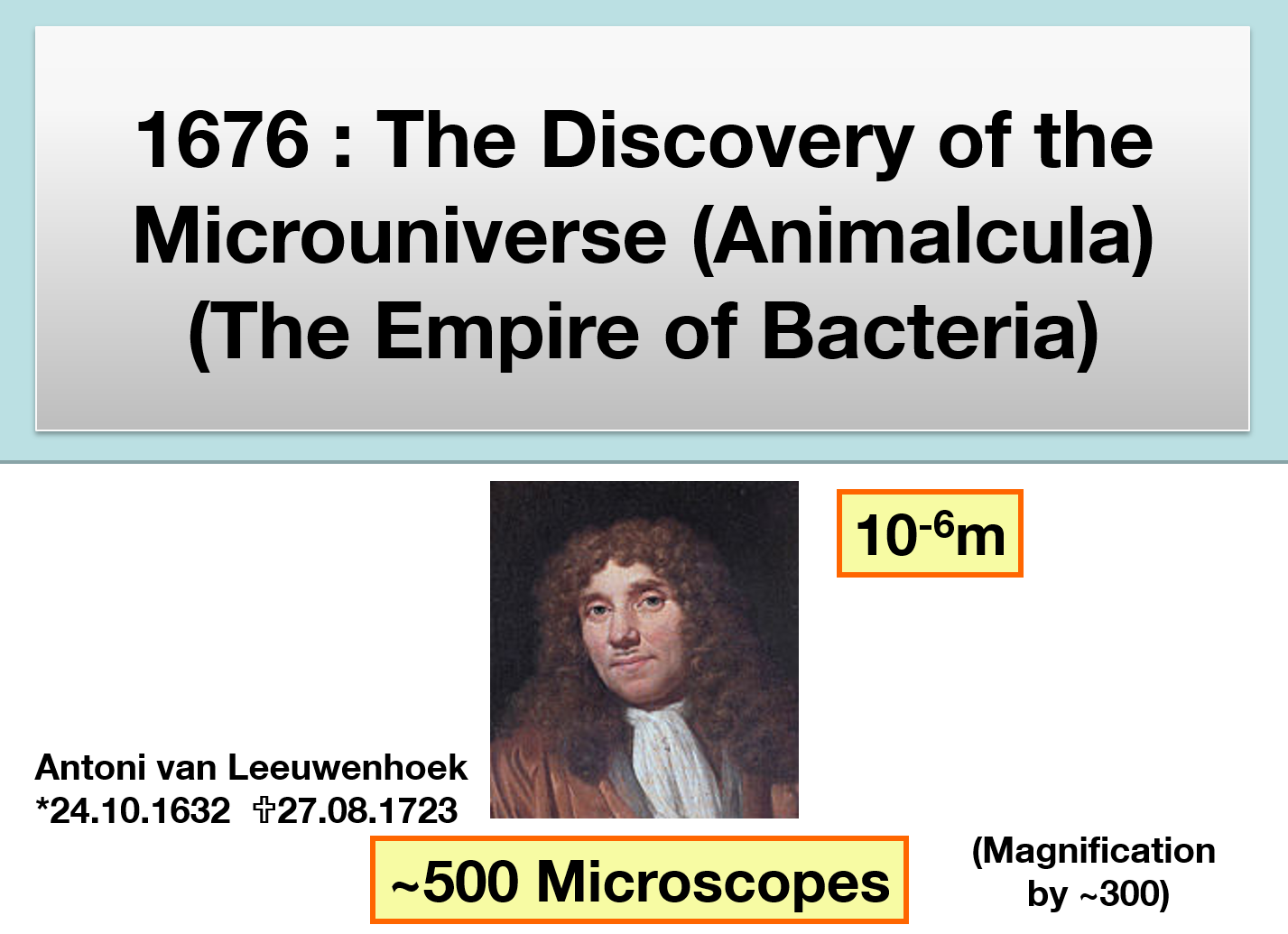}%
\caption{\it Antoni van Leeuwenhoek. (Wikipedia) \label{Antoni}}
\end{figure}

The year 2026 is rather special for those who study short distance scales. It marks the 350th anniversary of the discovery of the empire of bacteria made in 1676
by Antoni van Leeuvenhoek (1632-1723) (See Fig.~\ref{Antoni}).  
He called these small creatures {\it animalcula} (little animals). This 
discovery was a mile stone in our civilization  for several reasons.
 He discovered invisible to us creatures which over thousands of years 
 were systematcally killing the humans, often responsible for millions 
 of death in one year. While Antoni van Leeuvenhoek and his important follower Lazzaro Spallanzani (1729-1799) did not realize the danger coming from
 these new living species, fortunately 
L. Pasteur (1822-1895) and Robert Koch (1843-1910) as well as other {\it microbe hunters} not only realized 
 the danger coming from this tiny creatures but also developed weapons against
 this empire\footnote{A very interesting book by Paul de Kruif published 100 years ago, another anniversary, reports on this progress.}. Moreover, van Leeuvenhoek was the first human who looked at short distance scales invisible to  us, discovering thereby
 a new {\it underground world: microuniverse}. 

\begin{figure}[t]
\centering%
\includegraphics[width=0.8\textwidth]{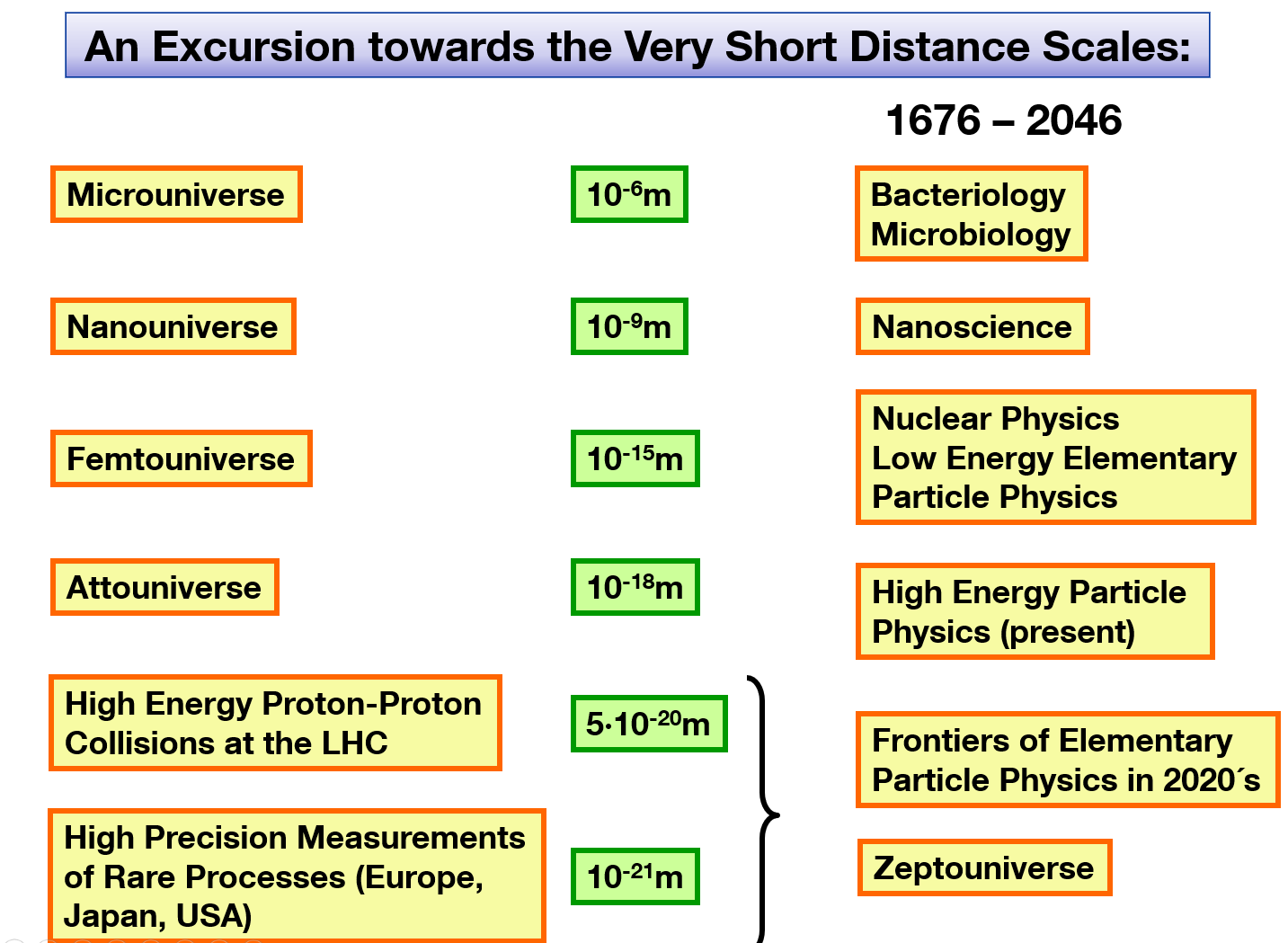}%
\caption{\it Progress in Resolving Short Distance Scales 1676-2026.\label{Progress}}
\end{figure}

While van Leeuvanhoek could reach the resolution down to roughly 
$10^{-6}$m, over the last  350 years this resolution could be improved 
 by many orders of magnitude reaching already in 1980 the resolution
 of $10^{-15}$m  in the context of low energy elementary particle physics. As seen in Fig.~\ref{Progress} it has further
 been improved by many orders of magnitude since then.

 On the way down to shortest distance 
scales scientists discovered {\it nanouniverse} ($10^{-9}$m), 
{\it femtouniverse}   ($10^{-15}$m) relevant for nuclear particle physics 
and low energy elementary particle physics and finally 
{\it attouniverse} ($10^{-18}$m)
that is the territory of contemporary high energy elementary particle physics.

In this decade and the coming decades we will be able to improve the resolution of  the short distance scales by at least
 an order of magnitude, extending the picture of fundamental physics 
down to scales $5\cdot 10^{-20}$m with the help of the high energy processes at the Large Hadron Collider (LHC). Further resolution, 
down to scales as short as $10^{-21}$m ({\it Zeptouniverse}) or even shorter scales,  should be possible with the help of 
high precision experiments in which flavour violating processes played  a 
prominent role for decades. This is evident from my recent {\em Flavour Autobiography}   \cite{Buras:2026vbp}.

\begin{figure}[t]
\centering%
\includegraphics[width=0.8\textwidth]{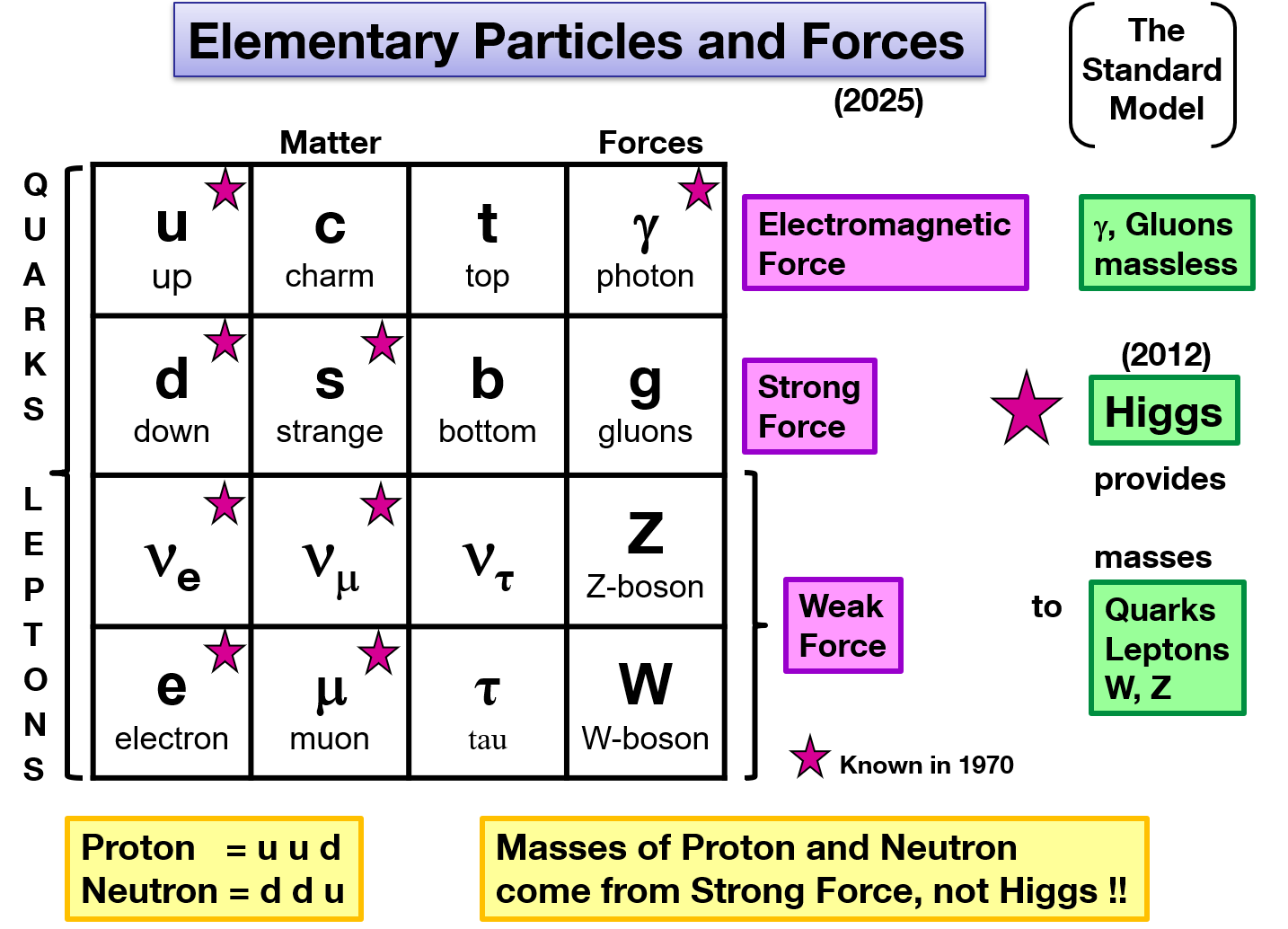}%
\caption{\it The Particles of the Standard Model.\label{SM}}
\end{figure}

Let us recall the particles of the Standard Model that all have
 been discovered. They are collected in Fig.~\ref{SM}. The small red stars  indicate which particles were known
 in 1970 when I was doing my master thesis. The big red star is reserved for the
 Higgs. Certainly a dramatic progress since 1970 has been made.

\begin{figure}[!bt]
\centering
\includegraphics[width = 0.75\textwidth]{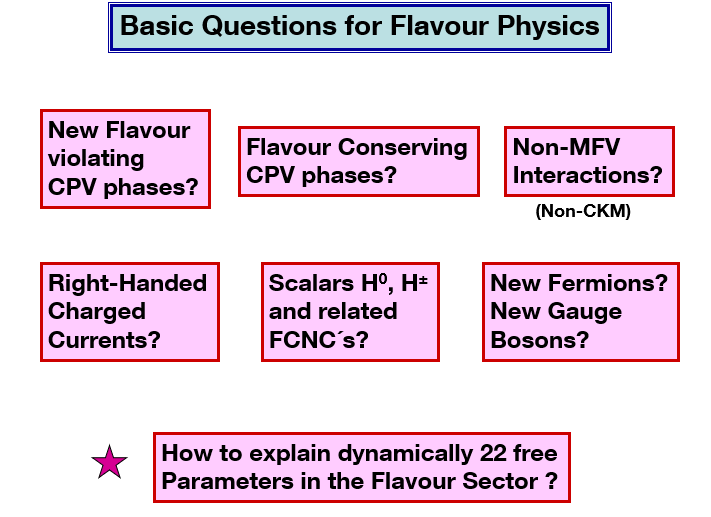}
 \caption{\it Basic Questions for Flavour Physics. }\label{BasicQ}
\end{figure}

My Flavour Autobiography is 323 pages long and I thought that it could turn
out to be useful to extract from it the most important strategies for the
search for new physics (NP). This is the main topic of this write-up. I arranged the material as follows.

In Section~\ref{NPQ} I will list the main reasons why we are sure that
NP beyond the SM must exist. In Section~\ref{DualP} I will present {\em Dual Picture of
  Short Distance Scales} that I have been presenting in numerous talks since 2009 but
not in print except for  \cite{Buras:2026vbp} recently. It demonstrates
in an impressive manner that the scales explored in particle physics
are indeed very short. In Section~\ref{BFramework} I will present an express summary of the basic framework in weak decays. To this end I will use 
two colourful figures from my talks. Having this I will list prerequisits for a successful search for NP and summarize their present status.

Next, in Section~\ref{WETSMEFT} I will describe very briefly the structure
of the Weak Effective Theory (WET) and of the Standard Model Effective Field
Theory (SMEFT). Begining with phenomenology in the SM I will describe in Section~\ref{BV}
the main ideas behind the strategy developed with Elena Venturini that avoids
present inconsistencies in the determination of the CKM element $\vcb$
that plays very important role in FCNC processes. Subsequently in Section~\ref{DNA} I will illustrate with colourful plots the so-called {\em DNA-Strategy}
developed this time with Jennifer Girrbach-Noe.

Next, in Section~\ref{NPMOD} a short description of most popular New Physics
models will be given. They can be distinguished through correlations between
various observables. We demonstrate this in Section~\ref{CORR} which
can be considered as an Album of correlations. It will consist of
several plots showing correlations in various NP models.
In Section~\ref{SMEFTCORR} we present the status of the analyses 
of correlations between observables within the SMEFT.
Finally in
Section~\ref{OUT} the summary and an outlook on the coming years will be given.

\section{Basic Questions beyond the Standard Model}\label{NPQ}

The Standard Model (SM) describes the data very well, but there are a number of fundamental questions that it cannot answer. They are well known, but let us list them again:

{\bf The Nature of Dark Matter:}
What is the particle nature of dark matter, which makes up about $27\%$  of the matter in the Universe but is absent from the SM?

{\bf Dark Energy and the Accelerated Expansion of the Universe:}
What is responsible for the observed accelerated expansion of the Universe?

{\bf Neutrino Masses:}
Why do neutrinos have tiny but non-zero masses, whereas they are massless in the SM? Are neutrinos Dirac fermions like in the SM or maybe Majorana fermions, i.e. their own antiparticles? This
    would indicate that lepton number is not a fundamental symmetry of nature.
    Neutrinoless double-beta decay would establish this property. Do right-handed neutrinos exist?

{\bf Matter–Antimatter Asymmetry:}
Why is the Universe dominated by matter rather than antimatter? This is related to the violation of CP-symmetry soon after the BIG BANG and is crucial for our existence. The size
    of CP violation in the SM is by far insufficient to explain this.

{\bf The Hierarchy Problem:}
Why is the Higgs boson mass so much lighter than the Planck scale? Is it a fundamental particle or a composite particle arising from some new strong dynamics responsible for spontaneous
    breakdown of electroweak symmetry of the SM? Is the form of Higgs potential affected by NP?

{\bf The Origin of Flavour:}
What determines the pattern of fermion masses and mixing angles?
Why is 
    the top quark mass larger by five orders of magnitude than the electron
    mass and eleven orders of magnitude larger than neutrino masses.
      What is the origin of the pattern of quark and lepton  interactions summarized by the
    CKM and PMNS matrices?
    
{\bf The Strong CP Problem:}
Why is CP violation in strong interactions so small?
Will axions arising from related solutions be discovered in the near future?

{\bf Unification of Forces:}
Do the fundamental forces unify at high energies? Numerous Ideas have been presented in the last fifty years.

{\bf Is Supersymmetry realized in Nature?:} While until the discovery of the Higgs supersymmetric extensions of the SM were leading our field, it does not appear to be the case now. But supersymmetry could still be realized at much higher
energy scales.

{\bf Quantum Gravity:}
How can gravity be consistently combined with quantum field theory?

{\bf Number of Generations:}
Why are there exactly three generations of quarks and leptons? While the
existence of a fourth generation of chiral fermions has by now been excluded, 
several generations of vector-like quarks and leptons  remain possible.

{\bf CP violation in the lepton sector ? :} If established, it would strengthen the case of leptogenesis related to the dominance of matter in the universe.
  
  In view of these questions it is obvious that new forces and new particles
  beyond those shown in Fig.~\ref{SM} must exist and our duty is to find them
  somehow. As far as flavour physics is concerned the basic questions are listed in more detail in Fig.~\ref{BasicQ}.

\section{Dual Picture of Short Distance Scales}\label{DualP}
Our colleagues in Astrophysics and Cosmology can easily impress the community
by showing impressive pictures of the Universe at large. One can look
at them with naked eyes or powerful telescopes. This is much harder in particle physics as with
good eyes one can sense $10^{-4}$ m and using microscopes, that are more powerful, one can see the  picture of nanouniverse but not really of short distance scales explored in particle physics.

While preparing my talk at the EPS-2009 in Cracow \cite{Buras:2009if}, I got the idea to improve
this situation by simply removing the minus sign from the exponential, that is
instead of using $10^{-x}$ m I proposed to use $10^{x}$ m. After all what matters
is the relative size to our hight which for adults is roughly $1.75\pm0.20$ m.

In fact in this manner one can really get impressed by the shortness of the scales
resolved in particle physics. In Figs~.\ref{Micro} and \ref{Zepto} I show the dual pictures of
the following Universes:

\begin{figure}[t]
\centering%
\includegraphics[width=0.45\textwidth]{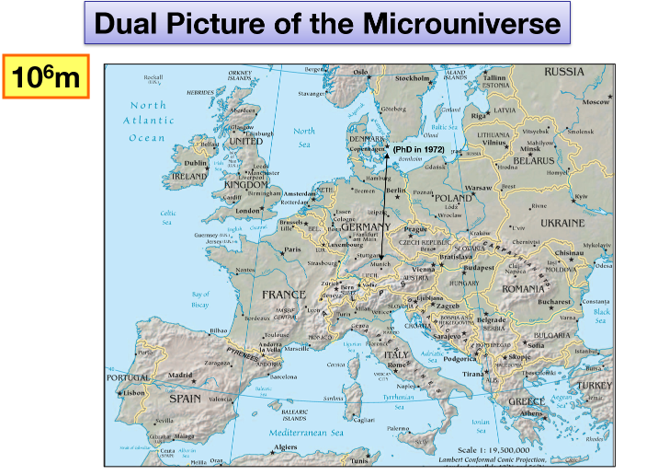}
\includegraphics[width=0.45\textwidth]{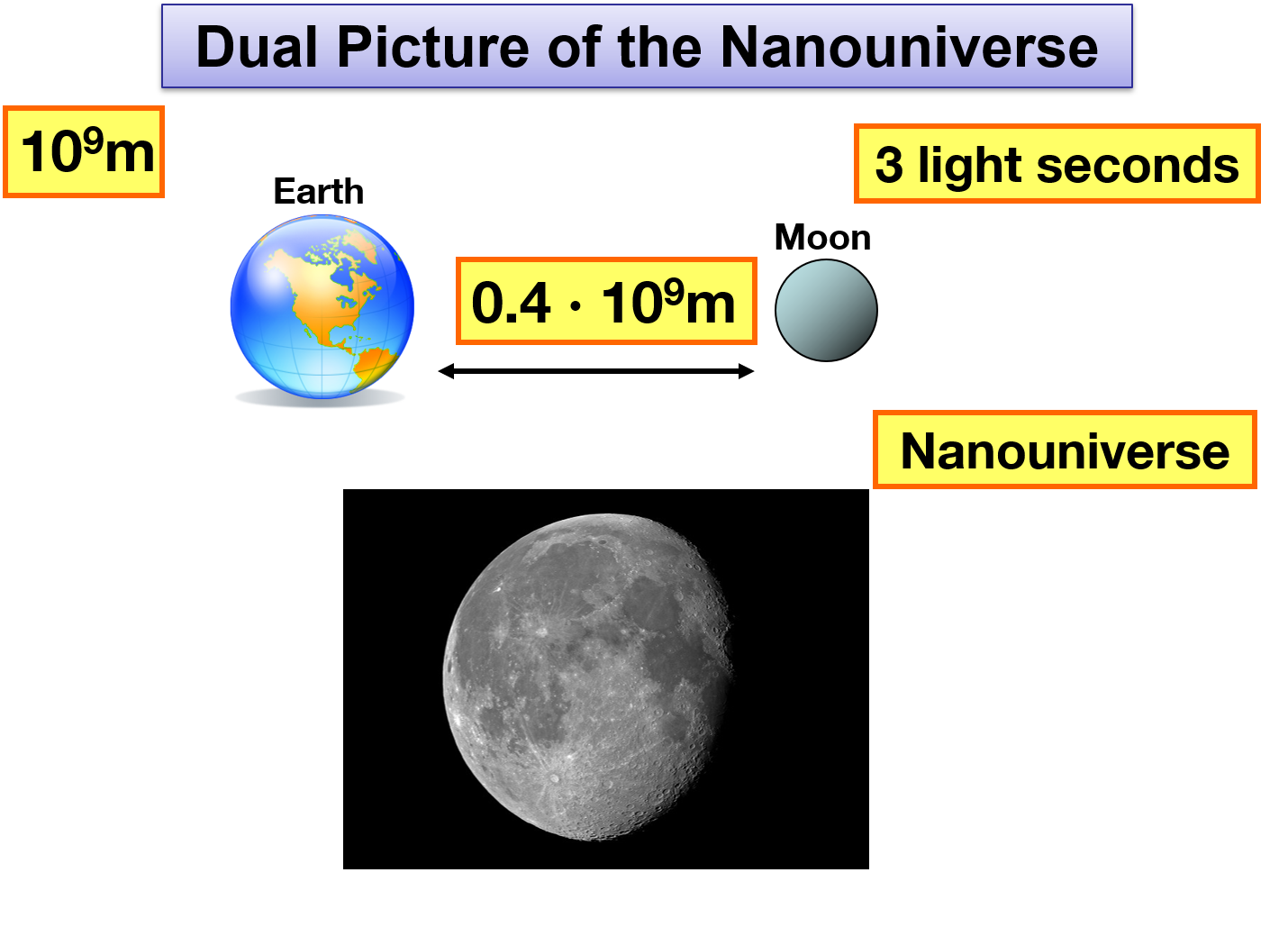}%
\caption{\it Dual Picture of the Microuniverse (Left) and of the Nanouniverse (right).\label{Micro}}
\end{figure}
\begin{figure}[t]
\centering%
\includegraphics[width=0.45\textwidth]{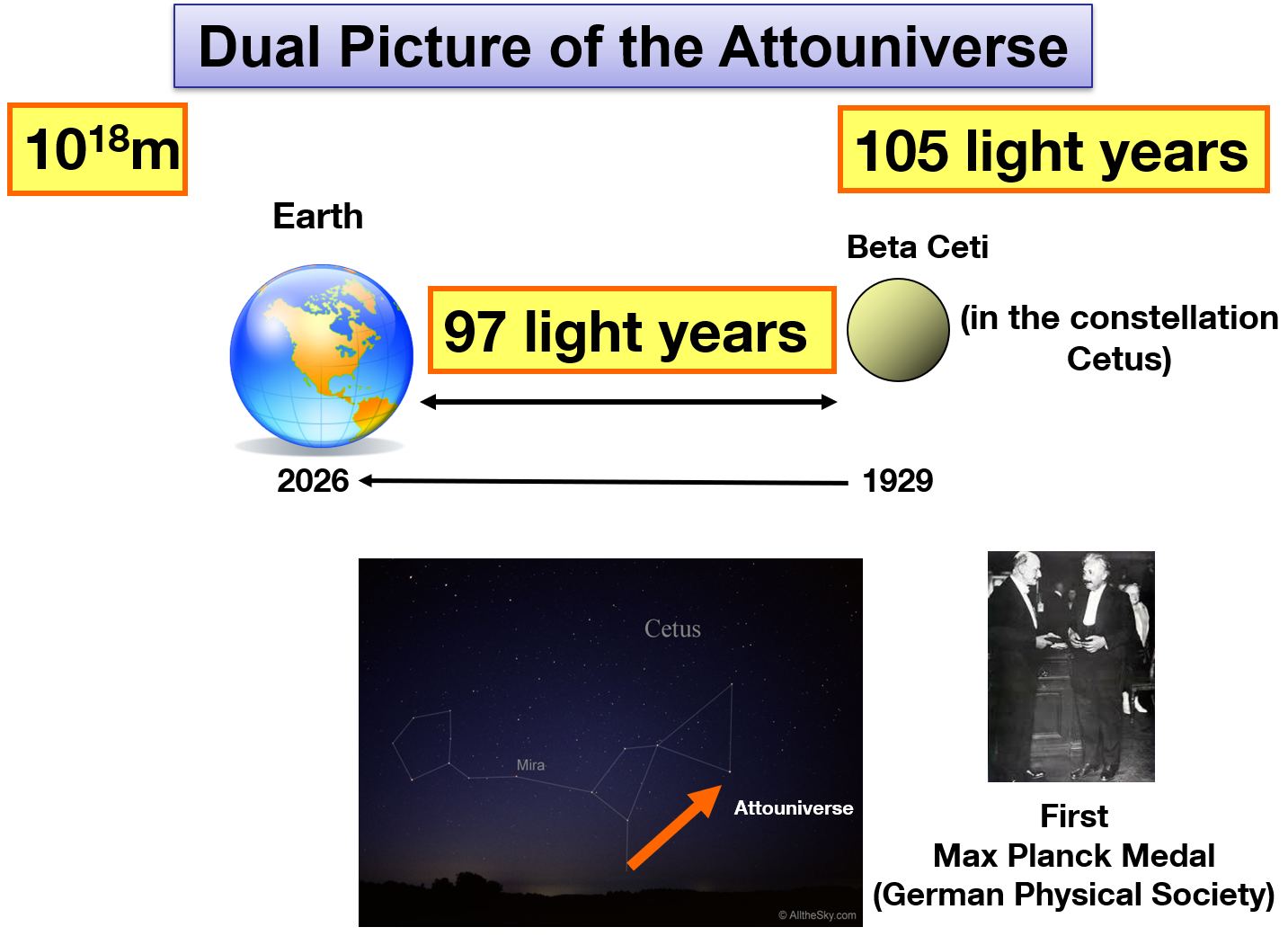}
\includegraphics[width=0.45\textwidth]{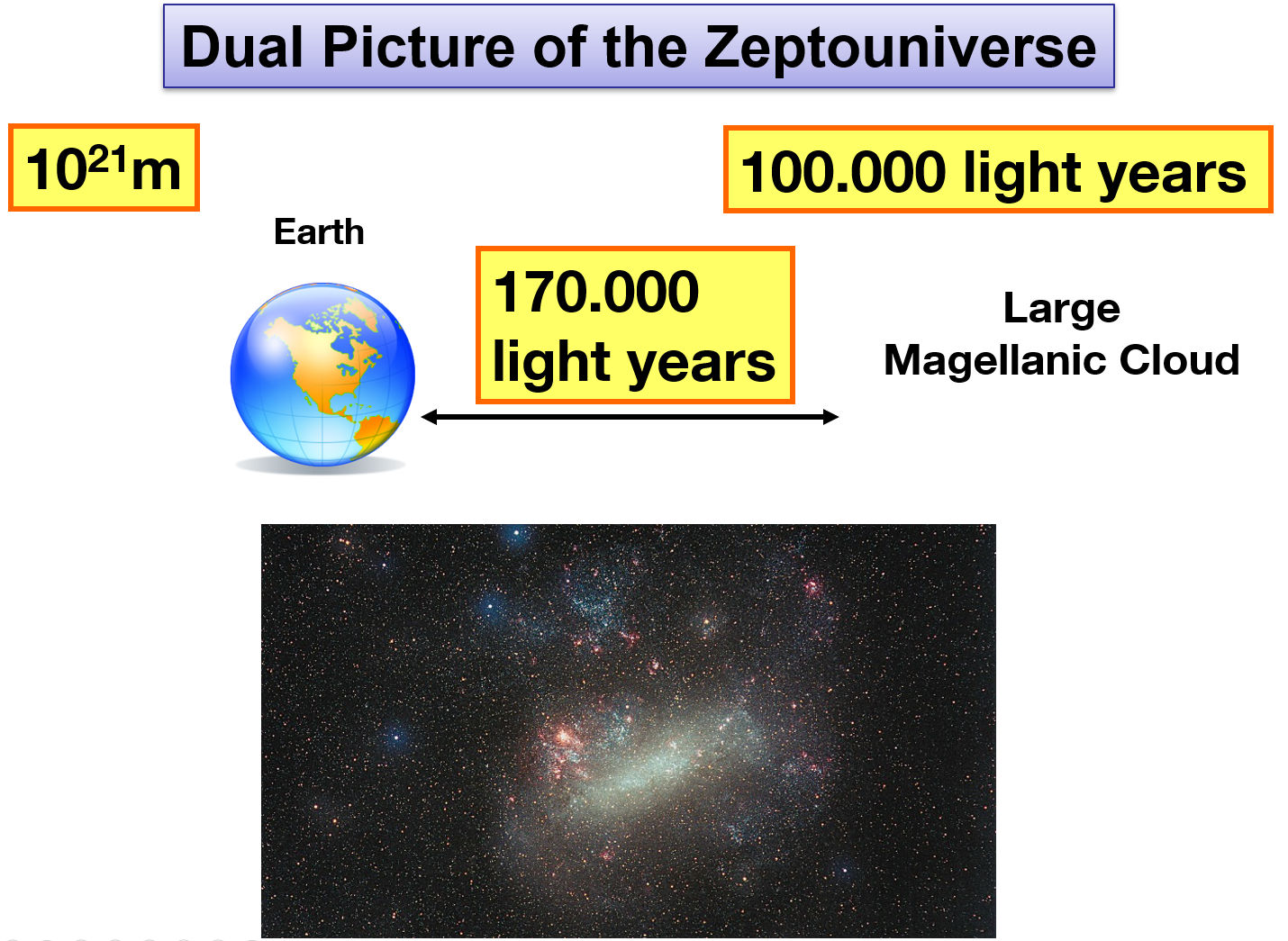}%
\caption{\it Dual Picture of the Attouniverse (left) and of the Zeptouniverse (right). (Wikipedia) \label{Zepto}}
\end{figure}
\begin{figure}[t]
\centering%
\includegraphics[width=0.45\textwidth]{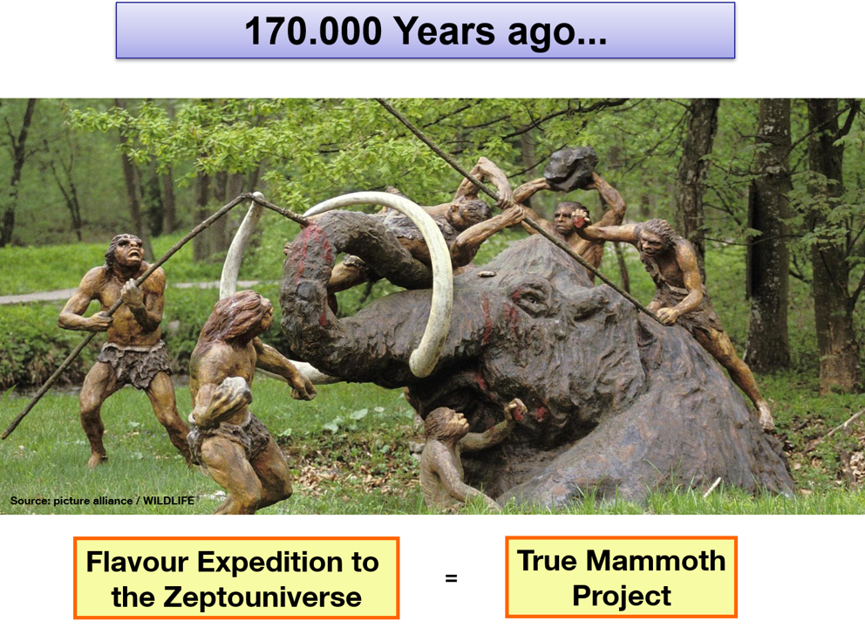}
\includegraphics[width=0.45\textwidth]{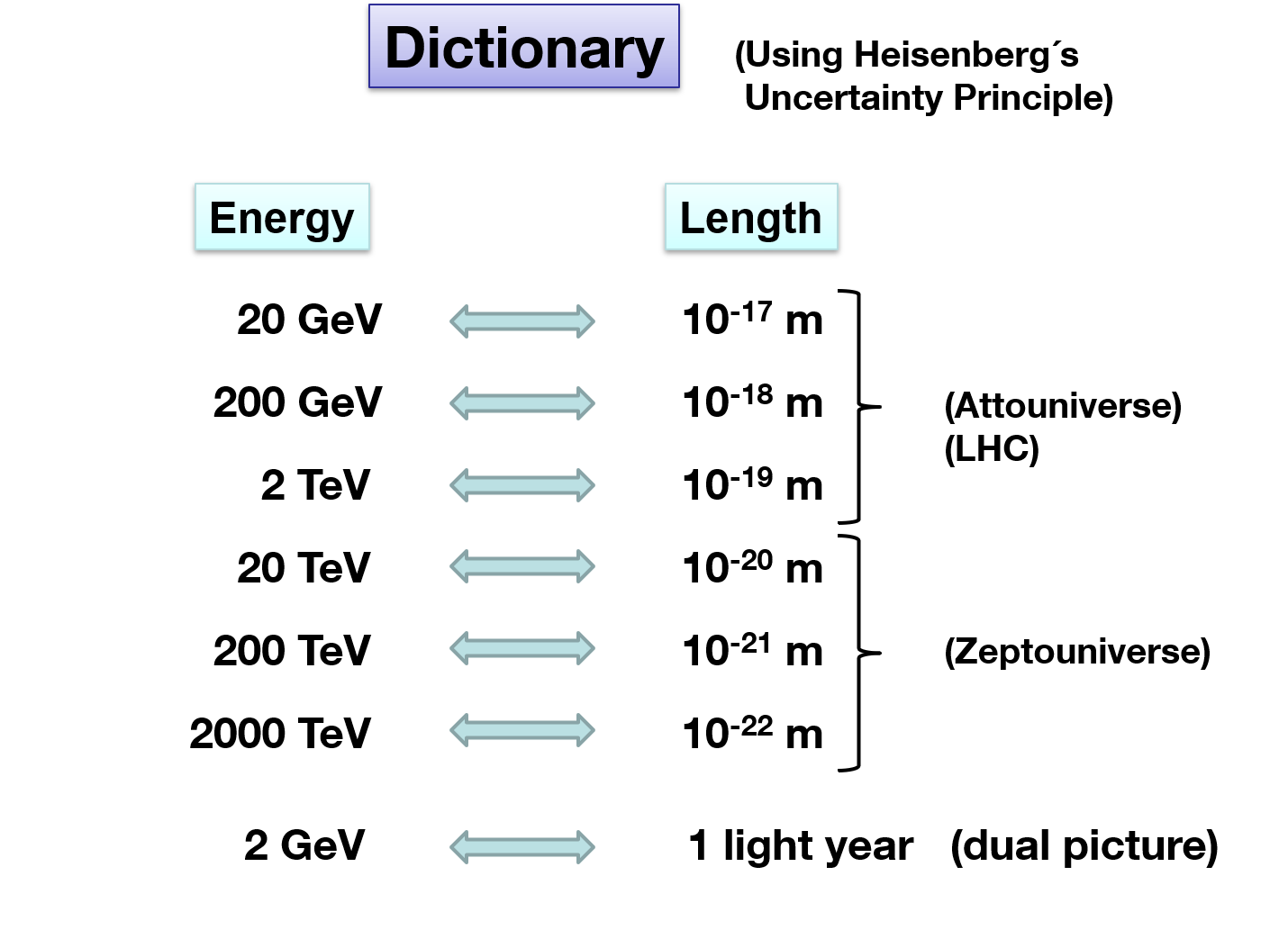}%
\caption{\it  Old Collaboration at Work (left) and the Dictionary (right).\label{Mamoth}}
\end{figure}

{\bf Dual Picture of the Microuniverse}

This corresponds to the distance from Munich to Copenhagen, where in 1972
I completed my PhD studies. See left picture in Fig.~\ref{Micro}.

{\bf Dual Picture of the Nanouniverse}

Looking at the moon one can get impressed by the distance scales explored
by the nanotechnologists. See right picture in Fig.~\ref{Micro}. Indeed, if we were as tall as the distance from
the Earth to the Moon, they would explore the distance scales of a few meters.
At first sight very impressive but cannot really compete with what particle
physicists can achieve. 

{\bf Dual Picture of the Attouniverse}

Indeed, already the Attouniverse is much more impressive than the Nanouniverse.
Let us assume that during the celebration of the first Max Planck Medal (1929)
given to Albert Einstein and Max Planck himself, a signal has been sent from
Beta Ceti (an object in the constellation Cetus, at a distance of roughly
$10^{18}$ m from us, see Fig.~\ref{Zepto}) (left) has been sent to the Earth. It will arrive only this year. 

{\bf Dual Picture of the Zeptouniverse}

We make next step to the Large Magellanic Cloud, see Fig.~\ref{Zepto} (right).
Let us then assume that
a signal from there will arrive on the earth this year. One can then ask what
was on the earth when this signal was sent to us. Certainly there was
no LHCb, CDF, ATLAS or Belle II collaboration but another one with different
goals as seen  on the left in Fig.~\ref{Mamoth}. The first guy on the left was probably the spokesman
of this collaboration. I doubt he expected to be cited by anybody in 2026.

As  is evident  from my  recent {\em Flavour Autobiography}   \cite{Buras:2026vbp} this picture fits very well to the efforts of many flavour physicists: 
{\em Flavour Expedition}  to very short distance scales with the goal to discover new animalcula. It  is indeed the true Mammoth Project. This will also be evident
as we continue.

The dictionary between energy scales explored by present and future colliders and distance scales used to construct
this dual picture of short distances is shown on the right in Fig.~\ref{Mamoth}.

 \section{Basic Framework and Prerequisits for Finding NP}\label{BFramework}

\begin{figure}[t]
\centering%
\includegraphics[width=0.8\textwidth]{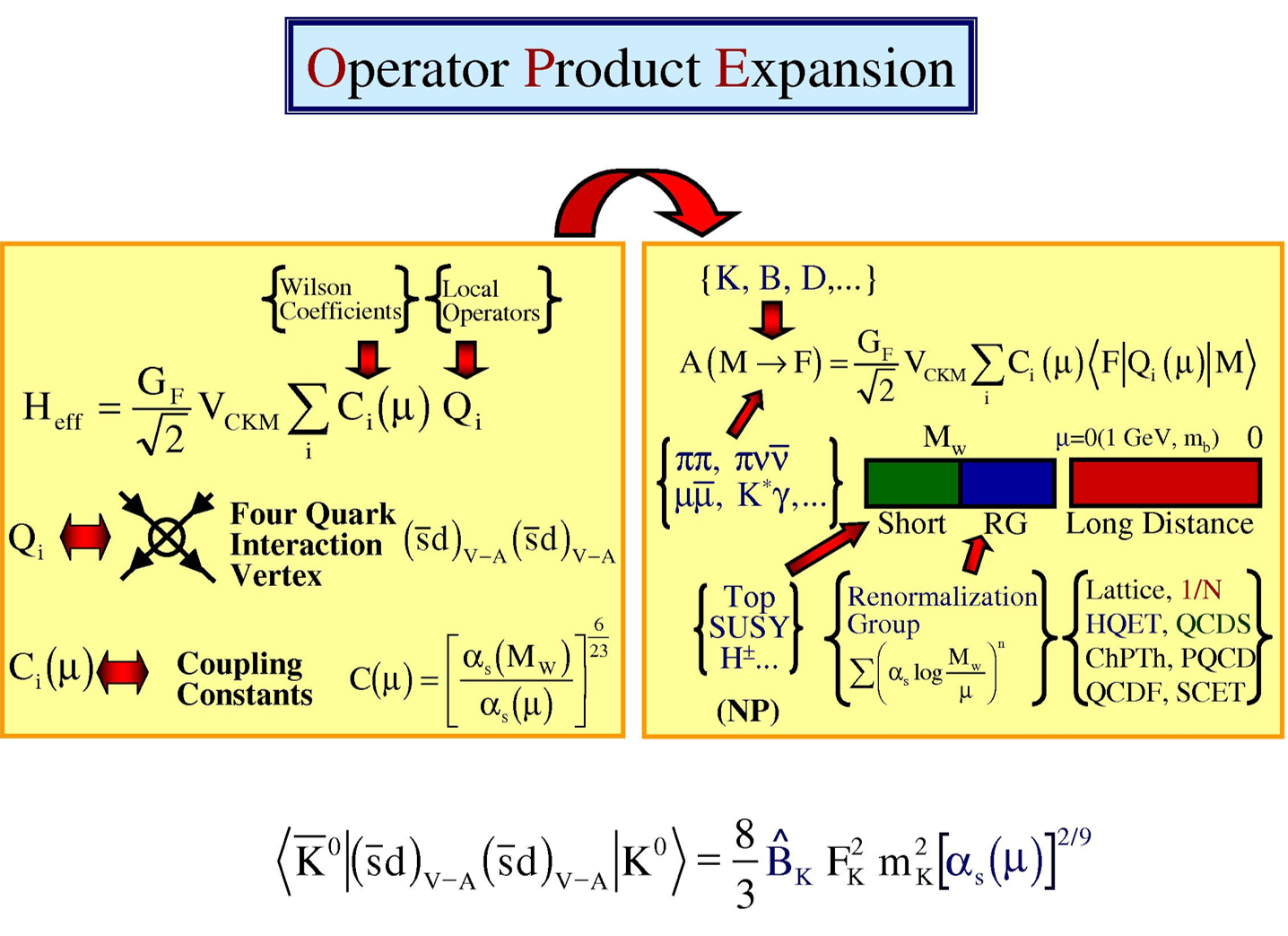}%
\caption{\it Operator Product Expansion. \label{OPE}}
\end{figure}

\begin{figure}[!tb]
\centering%
\includegraphics[width=0.8\textwidth]{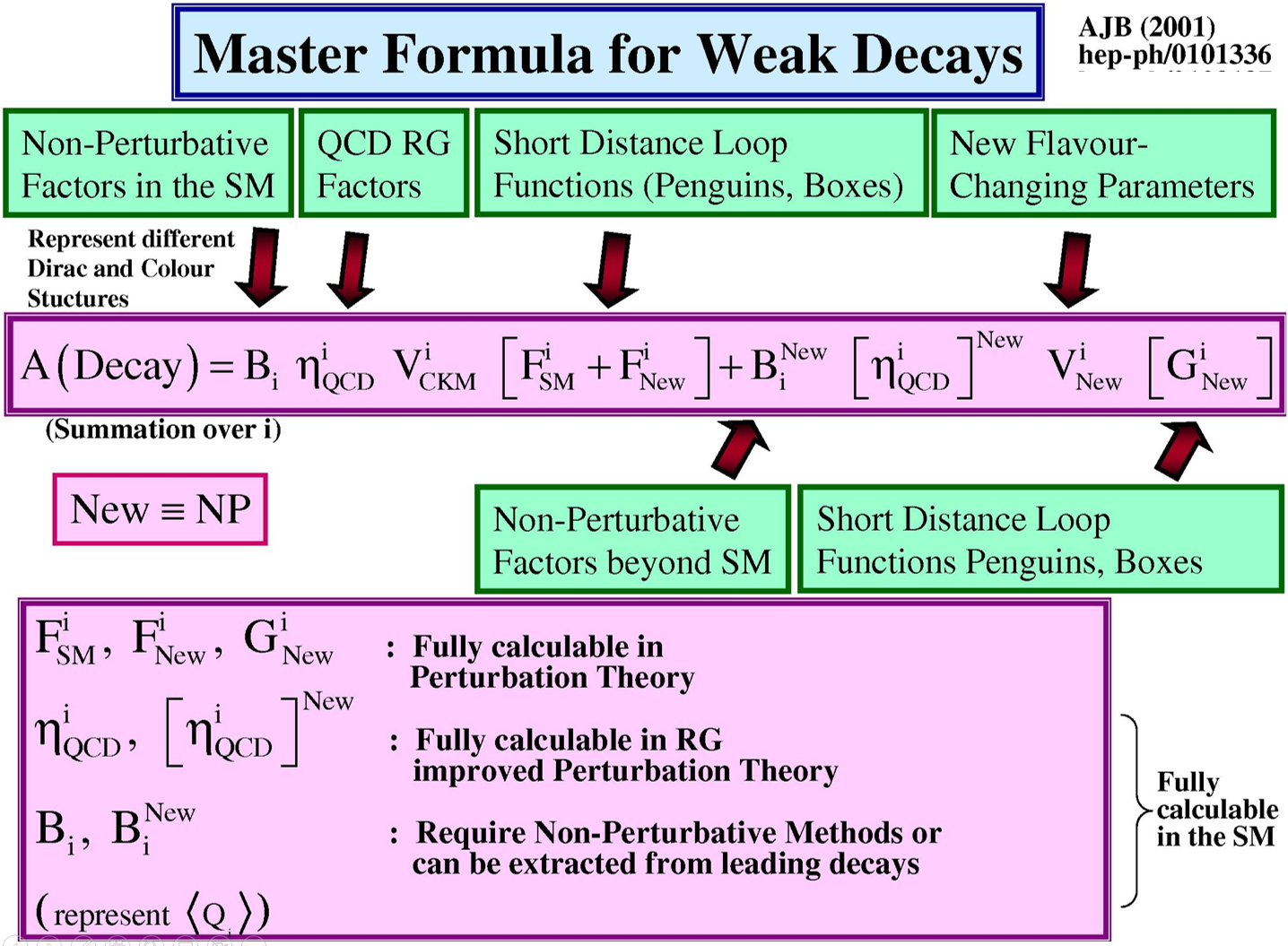}%
\caption{\it Master Formula for Weak Decay Amplitudes.\label{MASTERAP}}
\end{figure}

\begin{figure}[t]
\centering%
\includegraphics[width=0.8\textwidth]{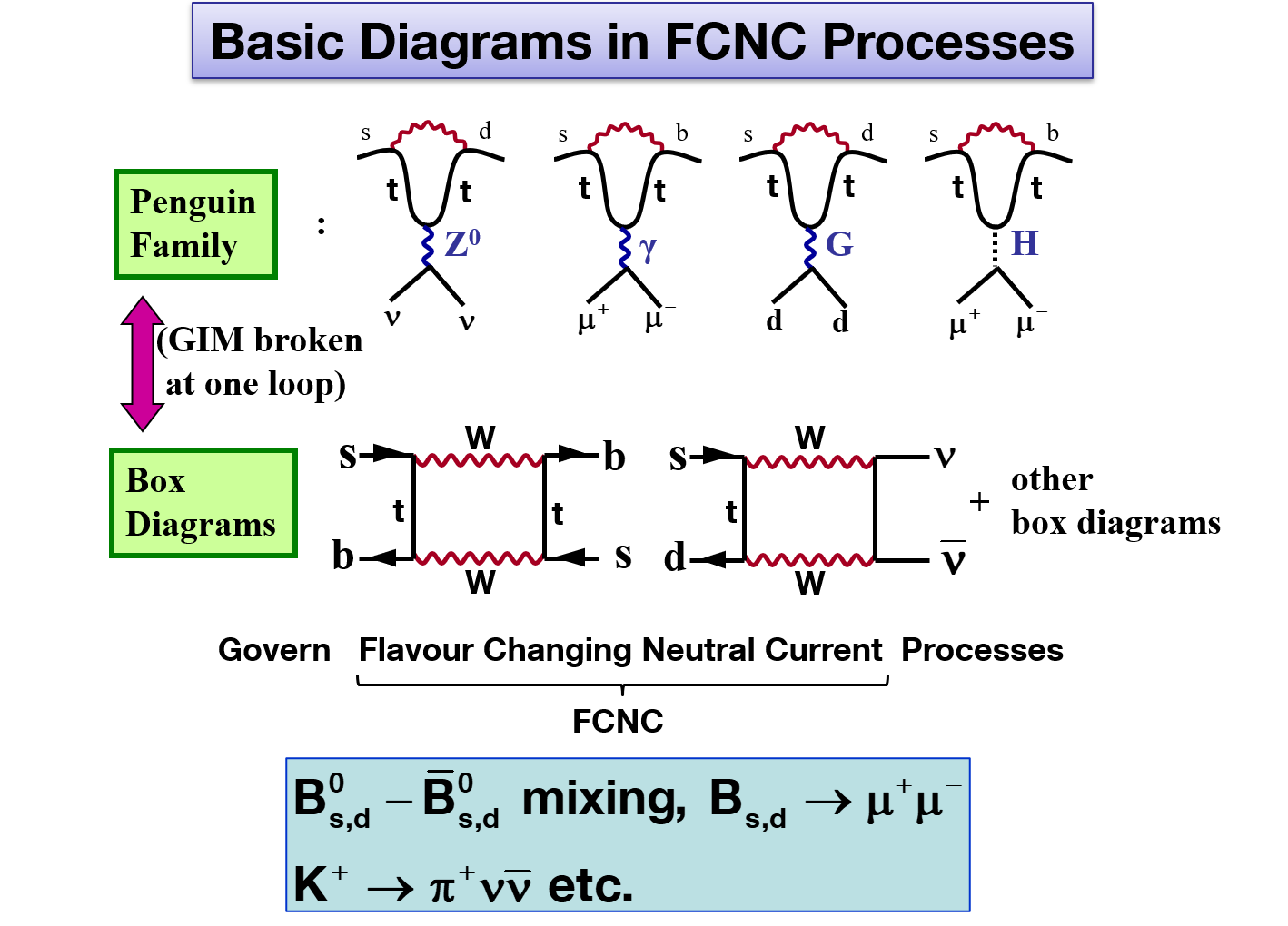}%
\caption{\it Basic Diagrams in FCNC Processes in the Standard Model.\label{Penguin}}
\end{figure}

  In Fig.~\ref{OPE} I show the basis of effective theories which is the operator product expansion, 
 the sum of operators multiplied by the Wilson Coefficients
 \cite{Wilson:1969zs,Zimmermann:1972tv,Wilson:1972ee}.   In Fig.~\ref{MASTERAP} the master formula for weak decay amplitudes is shown and
 different contributions are briefly explained. Detailed  explanations
 can be found in my book \cite{Buras:2020xsm} and in other reviews, in
 particular in my Les Houches lectures \cite{Buras:1998raa}.
The typical  Feynman diagrams in Flavour Physics of the SM are penguin and box diagrams. They are shown in Fig.~\ref{Penguin}.

This information should be sufficient for this lecture but 
the following paragraphs should bring additional insight in these matters.

%\section{Prerequisits: Measurements and Theory}\label{PR}
Let me stress first  that in the indirect search for NP with the help of
  weak decays, particle-antiparticle mixing observables, Electric Dipole Moments (EDMs) and other observables,  the first very important step is to obtain SM predictions as accurate as possible. Simply because  the first hints for NP in an indirect search will come from deviations from SM expectations.

  As the masses of the involved SM quarks and leptons, gauge bosons and of the Higgs
  are by now very well known and this also applies to gauge couplings, the
  main challenge in obtaining precise SM predictions for the processes in
  question are:
  \begin{itemize}
  \item
    Calculations of short distance QCD, QED and electroweak corrections. These are the loop functions $F^i_{\rm SM}$ and factors $\eta^i_{\rm QCD}$ in Fig.~\ref{MASTERAP}.
  \item
    Calculations of non-perturbative QCD effects. They are represented by the factors $B_i$ in Fig.~\ref{MASTERAP}.
  \item
    Determination of  CKM and PMNS parameters.
  \item
    Precise experimental measurements of the observables involved.
    \end{itemize}

  In my view the first challenge on this list has been overcome to a large extent as
  described in Part~V of \cite{Buras:2026vbp} and in \cite{Buras:2011we}.
Impressive progress has also been made
  by Lattice QCD (LQCD) in the evaluation of non-perturbative QCD effects with regular
  reports from FLAG \cite{FlavourLatticeAveragingGroupFLAG:2021npn}. This is in particular the case of particle-antiparticle mixing, weak decay constants,
  formfactors etc. Similar ChPT made important contributions here
\cite{Cirigliano:2011ny}.
In this
  manner leptonic  meson decays contain
  by now only small non-perturbative uncertainties. This is also the case
  of semileptonic FCNCs with neutrinos in the final state, where the main
  challenge are formfactors. Significant progress on them has been made
  in the last ten years \cite{Parrott:2022rgu,Parrott:2022zte,Parrott:2022smq,Khodjamirian:2023wol}.

  The case of semileptonic decays with charged
  leptons in the final state is another story because of non-factorizable
  long distance contributions that require detail studies. Presently
  they do not allow unique agreement between various theory groups on whether
  the anomalies observed in $b\to s \ell^+\ell^-$ transitions result
  from NP or non-perturbative effects. There is a very reach literature
  on this subject and I list here only few papers in which further references
  can be found \cite{Jager:2012uw,Jager:2014rwa,Gubernari:2020eft,Ciuchini:2022wbq,Gubernari:2022hxn,Gubernari:2023puw,Isidori:2024lng,Frezzotti:2025hif}.
    
  Unfortunately the situation is even worse in the case of
  non-leptonic meson decays where the non-perturbative uncertainties
  remain large. See Parts III and XI in \cite{Buras:2026vbp} for the most recent summary of the present status.

  Another issue are CKM parameters that are free parameters in the SM and
  have to be determined from experiment, preferably from tree-level decays
  that are believed to be less affected by NP than loop induced processes like
  FCNC processes within the SM. Unfortunately there are inconsistencies between
  different  determinations of some CKM parameters, in particular $\vcb$. This
  introduces significant uncertainties in decays like $\kpn$ and $\klpn$ that
  are sensitive functions of $\vcb$. I will return to this issue in Section~\ref{BV}.

  Finally, crucial are the measurements of various observables, like weak decay branching
  ratios, CP asymmetries, mass differences $\Delta M_{s,d}$ etc. Some are
  already measured with respectable precision, other should be measured in the coming years
  but for several we will have to wait still for a decade or more. The related
  branching ratios are simply tiny.

  \section{SMEFT and WET}\label{WETSMEFT}
\begin{figure}[t]
\centering%
\includegraphics[width=0.8\textwidth]{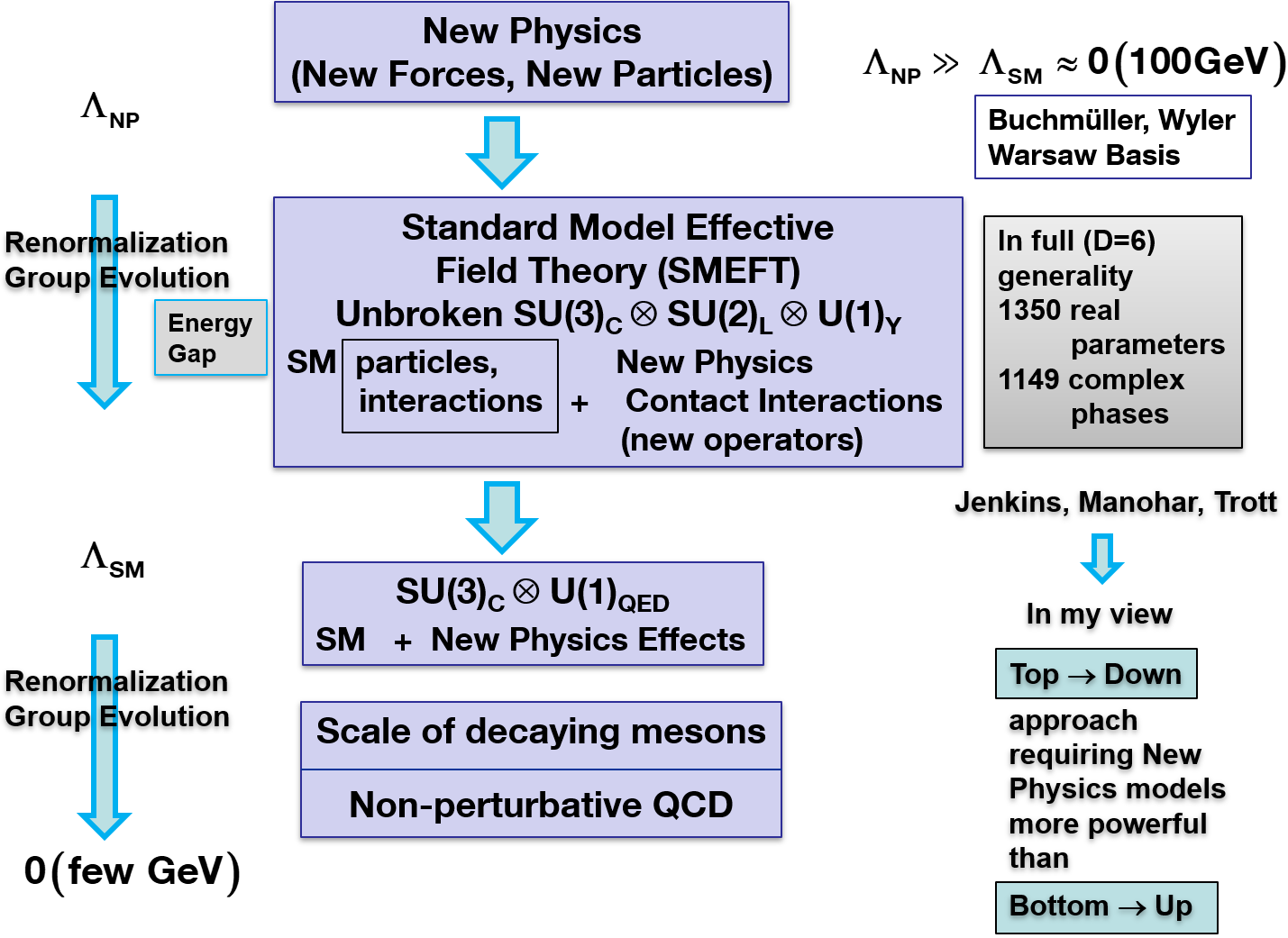}%
\caption{\it Grand View of the RG Evolution in the SMEFT.\label{GVSMEFT}}
\end{figure}

    These days two general effective theories play an important role in particle
  physics: SMEFT and WET.  I want to describe briefly what they are. In this context Fig.~\ref{GVSMEFT} is very useful. It deals with the so-called {\em Top-Down Approach}.

  \subsection{SMEFT}
Let us assume we constructed  a new model, the UV completion of the SM, which is supposed to answer some of the questions listed above and to describe the anomalies (departures from SM predictions) observed in experiments. It is an extension of the SM  that contains new particles and new forces. The latter are described by new gauge symmetries that are spontaneously broken directly or in steps down to the SM gauge group at some scale $\Lambda$ at which the lightest new particles have been integrated out. Below $\Lambda$ only SM particles appear and the gauge symmetry is the unbroken gauge group of the SM,
\be\label{SMGAUGE}
\text{SU(3)}_C\otimes\text{SU(2)}_L\otimes\text{U(1)}_Y\,,
\ee
provided $\Lambda$ is significantly larger than the electroweak (EW) scale $\muEW$.

The Effective Field Theory (EFT) with the unbroken SM gauge group  is usually called the Standard Model Effective Field Theory (SMEFT) \cite{Buchmuller:1985jz,Grzadkowski:2010es} because at low-energies, it should reduce to the SM, provided no undiscovered weakly coupled {\em light} particles exist, like axions or sterile neutrinos. For recent reviews on the SMEFT see  \cite{Brivio:2017vri,Isidori:2023pyp,Aebischer:2025qhh}.

However, the presence of new operators, in particular dimension six ones, whose Wilson coefficients (WCs) can be calculated in a given model as functions of its parameters, can introduce very significant modifications of SM predictions for flavour observables. Also the values of SM parameters can be modified in this manner. This is also the case for the WCs of the involved operators which are generated in the SM and, being modified, contain information about NP beyond the SM. Moreover there is a multitude of operators that are strongly suppressed in the SM but can be important in the SMEFT. Therefore the name SMEFT is to some extent  a misnomer but as the basic gauge symmetry is the SM one, it has been accepted by the particle physics community.

The structure of the interactions in the SMEFT, even if governed by the SM gauge group,  depends on the UV completion considered. This is important, because otherwise there would be no chance to find out anything about the dynamics above the scale $\Lambda$. In particular the number of free parameters will depend on the fundamental theory or model considered. These parameters can be conveniently defined at the scale $\Lambda$. Their number depends on the theory considered and it will not change when the renormalization group (RG) evolution down to the electroweak scale is performed. Only their values will change as well as the values of the WCs in the effective theory describing the physics between the scale $\Lambda$ and the electroweak scale.

This physics, beyond the one of the SM, is described by an effective Lagrangian
\begin{equation}\label{BASICNP}
\mathcal L^{(6)} =   \sum_{k} C_k^{(6)} Q_k^{(6)}\,,
\end{equation}
where for simplicity we keep only the contributions of dimension six. Other
contributions are discussed in details in numerous papers like in
\cite{Brivio:2017vri,Isidori:2023pyp,Aebischer:2025qhh} and the ones listed
below.

The number of operators and their Dirac structure depends generally on the model considered. However, to be prepared for all possible models and to develop a general framework one can classify operators in full generality by imposing only the SM gauge symmetry in (\ref{SMGAUGE}). This has been done in~\cite{Buchmuller:1985jz,Grzadkowski:2010es}. The second of these papers removed certain redundant operators present in the first one and these days  the results of~\cite{Grzadkowski:2010es} are used. However, by no means the pioneering work of
Buchm\"uler and Wyler should be underestimated.
The corresponding RG analysis at leading order (LO) of all these operators has been presented in~\cite{Jenkins:2013zja,Jenkins:2013wua,Alonso:2013hga}.

It turns out that for three generations of fermions, there  are 2499 independent operators $ Q_k^{(6)}$ (59 irreducible flavour representations) that do not violate baryon and lepton number. This means that the RG analysis involves in full generality a $2499\times 2499$ anomalous dimension matrix\footnote{Fortunately, as seen in Fig.~1 of \cite{Aebischer:2025qhh} there very many vanishing elements in this matrix.} with the evolution governed by the Higgs self-coupling $\lambda$~\cite{Jenkins:2013zja}, {the} Yukawa couplings~\cite{Jenkins:2013wua} and the SM gauge interactions~\cite{Alonso:2013hga}. The status of the calculations of anomalous dimesions at the leading and next-leading order is summarized in \cite{Buras:2026vbp,Aebischer:2025qhh} with
the second paper presenting technical details.

\subsection{WET}
The SMEFT Renormalization Group Equations (RGEs) allow calculating the WCs of the SMEFT at the electroweak scale. Below the electroweak scale the interactions are governed not by the full gauge symmetry of the SM but by
\be\label{SMWET}
\text{SU(3)}_C\otimes\text{U(1)}_{\text{QED}}\,,
\ee
which is very familiar to us. However, in the presence of NP there are new operators generated already at the NP scale and/or at the electroweak scale through RG running in the SMEFT and through the matching onto the low-energy theory described by this reduced symmetry. Consequently, the starting point for the RG evolution from the electroweak scale down to the low energy scale differs from the one encountered within the SM. Not only the initial conditions for SM WCs at the electroweak scale can be modified by NP but also new WCs might be present. The resulting effective theory below the electroweak scale, although as the SM based on the gauge symmetry of QCD~$\times$~QED, differs then from the SM and is called the Week Effective Theory (WET). It should be stressed that this theory does not involve $W$, $Z$ gauge bosons, the Higgs boson and the top quark as dynamical degrees of freedom. Moreover, there are no elementary scalars in this theory.

\begin{figure}[t]
\centering%
\includegraphics[width=0.90\textwidth]{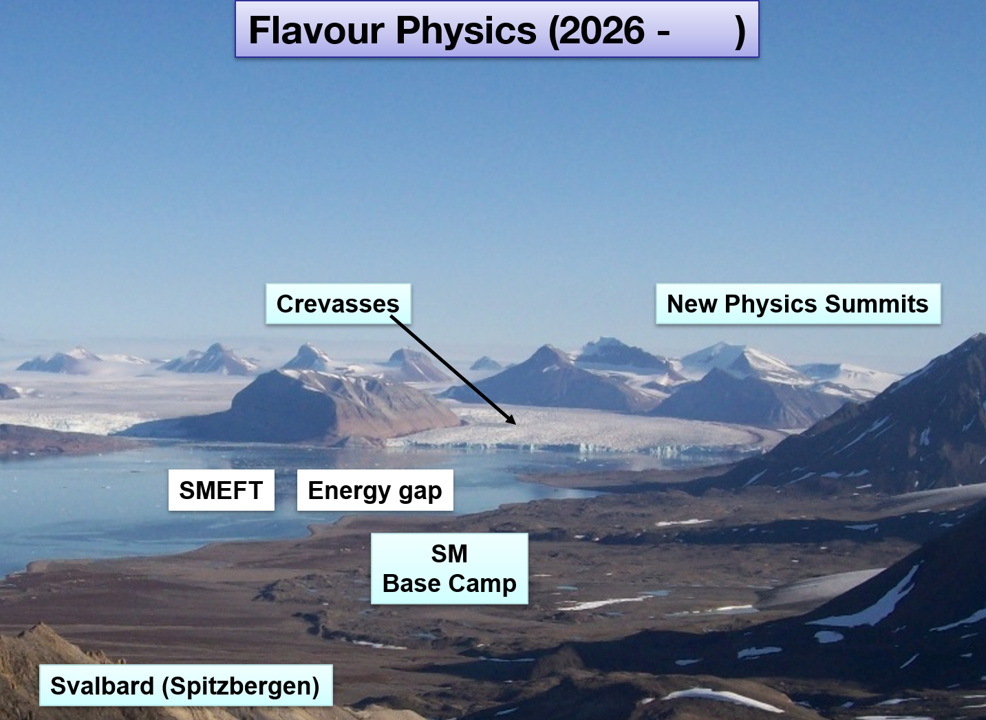}%
\caption{\it SMEFT and New Physics Summits.\label{Svalbard}}
\end{figure}

The counting of all possible gauge-invariant operators of the WET has been accomplished in~\cite{Jenkins:2017jig}. There are 70 
hermitian $D=5$ and 3631 hermitian $D=6$ operators that conserve baryon and lepton numbers, as well as $\Delta B=\pm \Delta L=\pm 1$, $\Delta L=\pm 2$, and $\Delta L=\pm 4$ operators. Among the 3631 operators in question 1933 are CP-even and 1698 CP-odd. 

This counting shows clearly that in order for a model to be predictive the number of free parameters has to be reduced by much. In particular flavour symmetries play here an important role. The reviews in \cite{Isidori:2023pyp,Aebischer:2025qhh} describe them in some detail and contain several references to original papers in which these flavour symmetries have been proposed and investigated phenomenologically.

It is now evident that searching for NP indirectly is a  true Mammoth Project.
This becomes even clearer by reading Part II of \cite{Aebischer:2025qhh} which
makes the anatomy of operator mixing in the process of the renormalization group evolution from $\Lambda$ down to $\muEW$.
But we should not give up because NP could be as beautiful as the photo in
Fig.~\ref{Svalbard} which has been taken by our younger son Allan during his expedition related to climate rather than flavour.

This photo describes actually so-called  Bottom-Up Approach. In this approach, one first tries to describe various departures of the low energy data from SM expectations by modifying the WCs of SM operators and/or adding new operators. Sometimes only a small number of new operators and related WCs is sufficient to describe the data. Having their values at the electroweak scale one can then as a first step construct simplified NP models which could imply the presence of these new operators and the modifications of SM WCs. Subsequently one can try to develop a more sophisticated model like in the top-down approach from which these WCs would result.

Thus in practice one tries to borrow the lessons from both approaches because while by definition in the top-down approach one knows better the fundamental new theory at short distance scales, in the bottom-up approach one has close contact with the data which the fundamental theory is supposed to describe.

But the route can be very tough as summarized in Fig.~\ref{Svalbard} with the
crevasses representing difficult  theoretical calculations and difficult  experimental measurements.

Let us then move to a strategy which allows efficiently to identify anomalies in experimental data giving signs of the NP at work.

\boldmath
\section{Removing $\vcb$ Uncertainties}\label{BV}
\unboldmath
As already stressed above, in order to identify the presence of NP high
precision of SM prediction is crucial. This is the main goal of the strategy
described now.
In order to motivate this strategy in explicit terms, let us recall  the values of $\vcb$
extracted from inclusive and exclusive tree-level semi-leptonic $b\to c$ decays
\cite{Finauri:2023kte,FlavourLatticeAveragingGroupFLAG:2021npn} 
\be
\vcb_\text{incl}=(41.97\pm0.48)\cdot 10^{-3},\qquad \vcb_\text{excl}=(39.21\pm0.62)\cdot 10^{-3}\,.
\ee
As rare K and B decays and mixing parameters are sensitive functions of $\vcb$, varying it from $39\cdot 10^{-3}$ to $43\cdot 10^{-3}$ changes $\Delta M_{s,d}$ and $B$-decay branching ratios
by roughly $21\%$, $\kpn$ branching ratio by $31\%$, $\varepsilon_K$ by $39\%$
and $\klpn$ and $K_S\to\mu^+\mu^-$ branching ratios by $48\%$.

Based on my report in Part~V in \cite{Buras:2026vbp},
these uncertainties are clearly a disaster for those like me, my collaborators
and other experts in NLO and NNLO calculations who spent decades to reduce theoretical uncertainties in basically all important rare $K$ and $B$ decays and
quark mixing observables down to $(1-2)\%$.

It is also a disaster for lattice QCD experts who for quark mixing observables and in particular  meson weak decay constants achieved the accuracy at the level of a few percent. Also significant progress on formfactors from LQCD and Light-Cone Sum Rules has been done.

 \begin{figure}
\centering
\includegraphics[width = 0.55\textwidth]{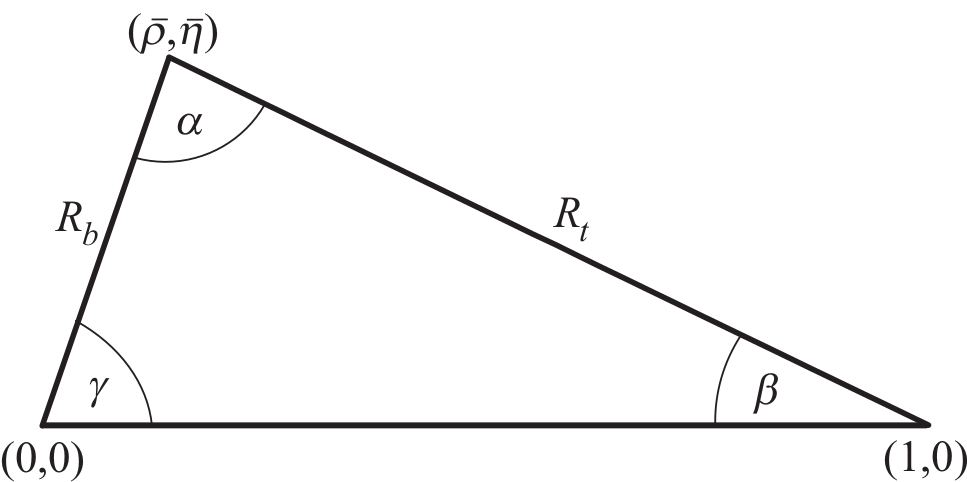}
 \caption{\it The Unitarity Triangle. }\label{UUTa}
\end{figure}

\begin{figure}[t!]
\centering%
\includegraphics[width=0.48\textwidth]{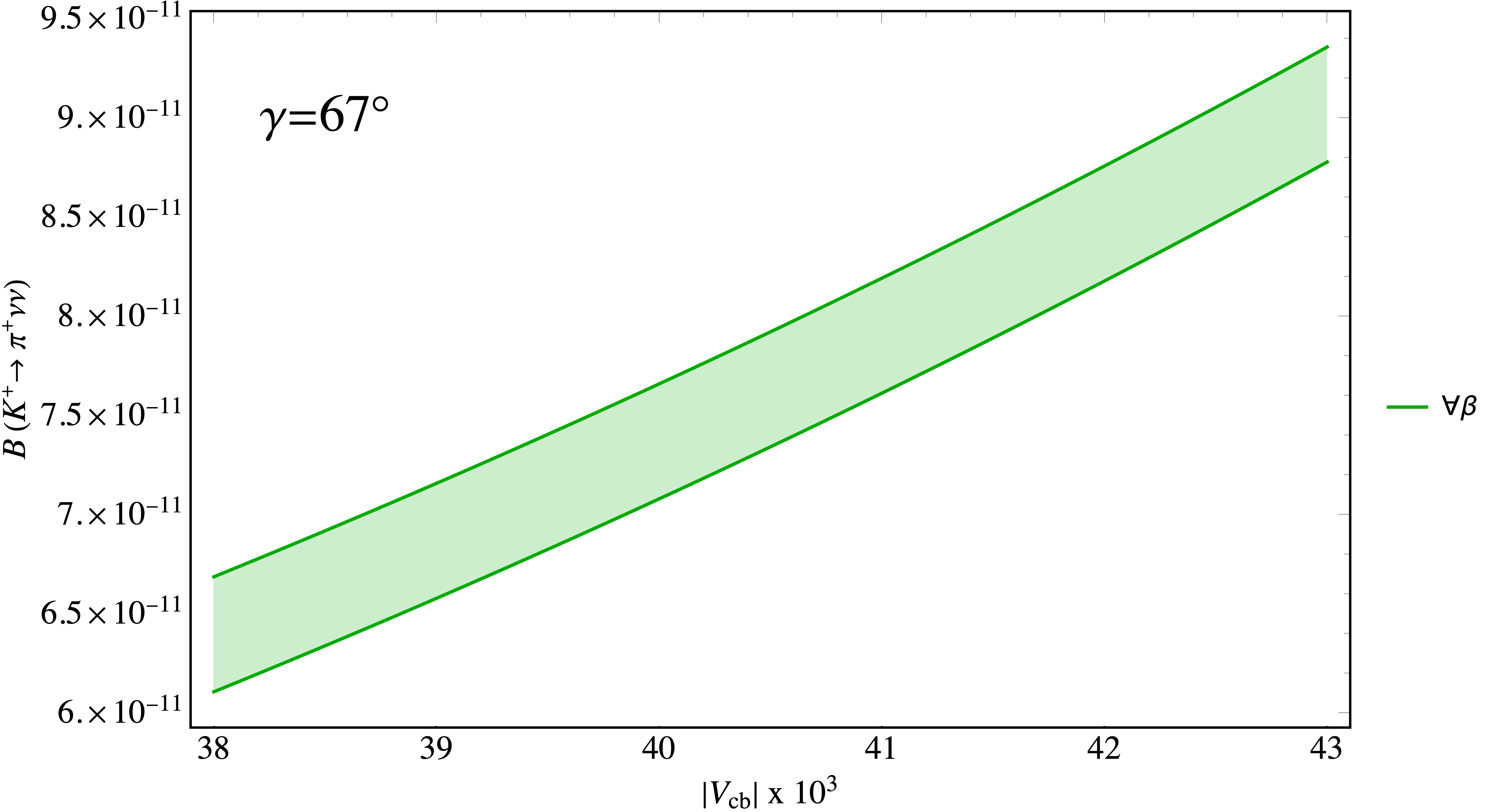}%
\hfill%
\includegraphics[width=0.48\textwidth]{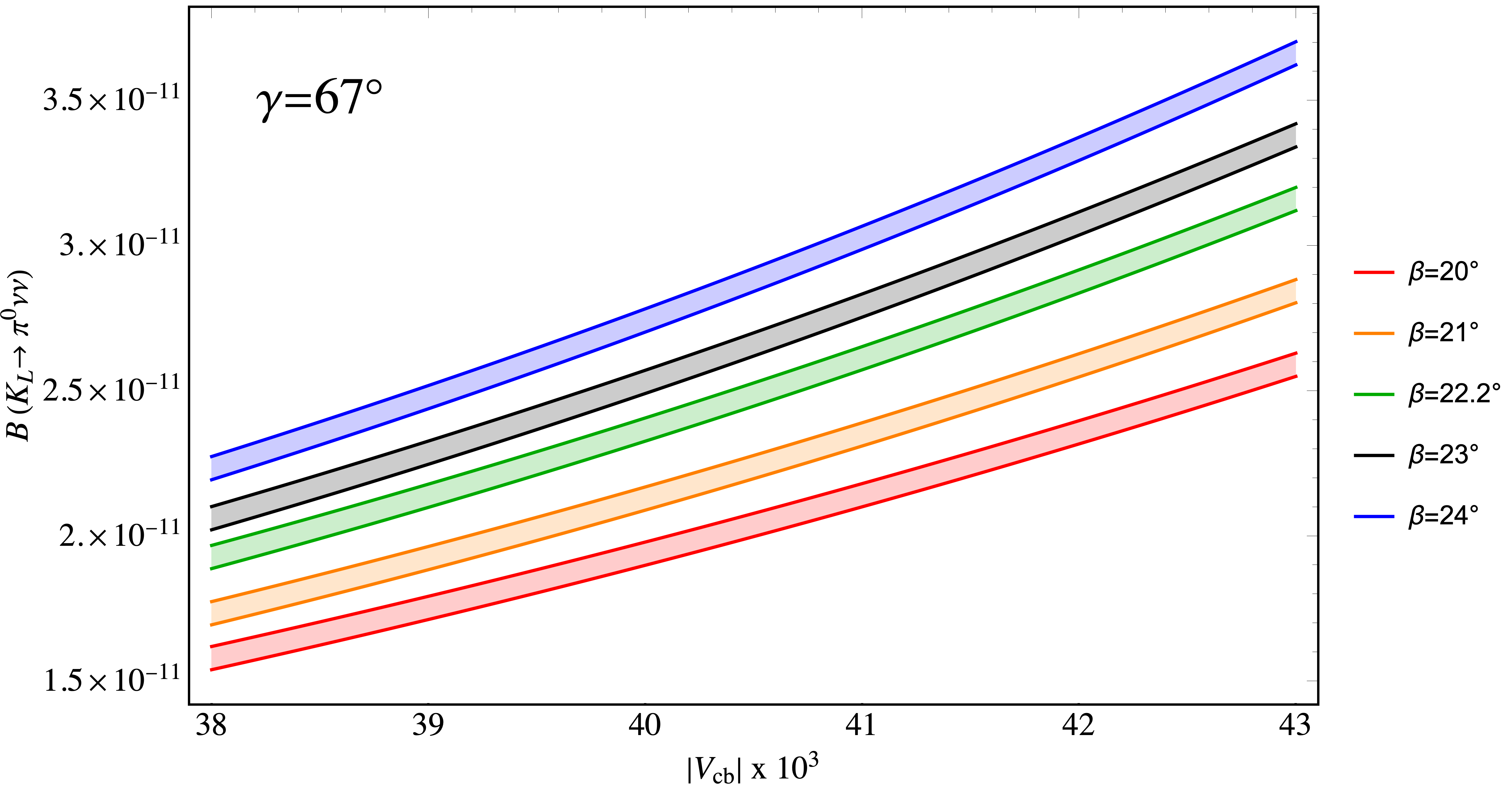}\\
\includegraphics[width=0.48\textwidth]{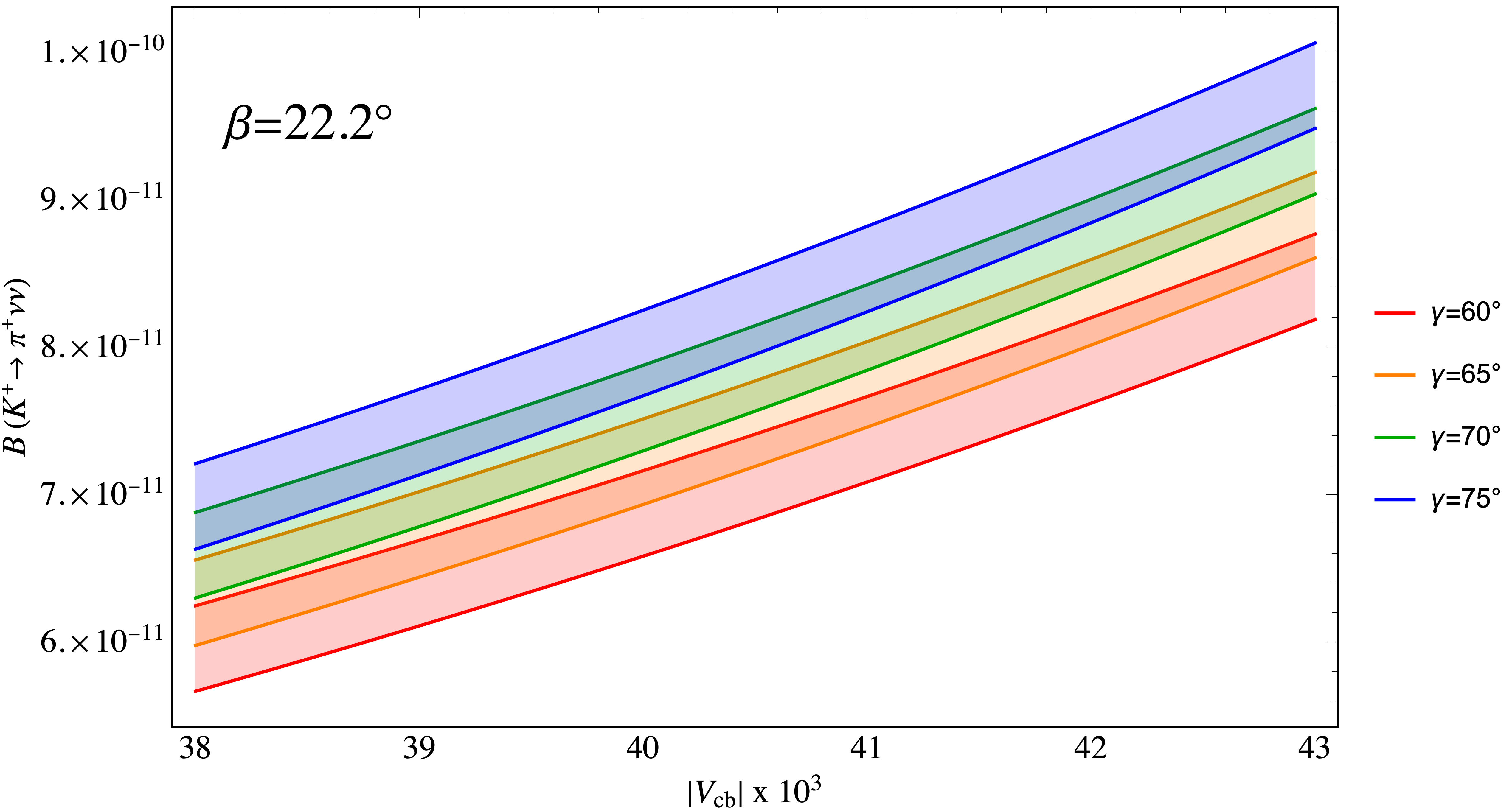}%
\hfill%
\includegraphics[width=0.48\textwidth]{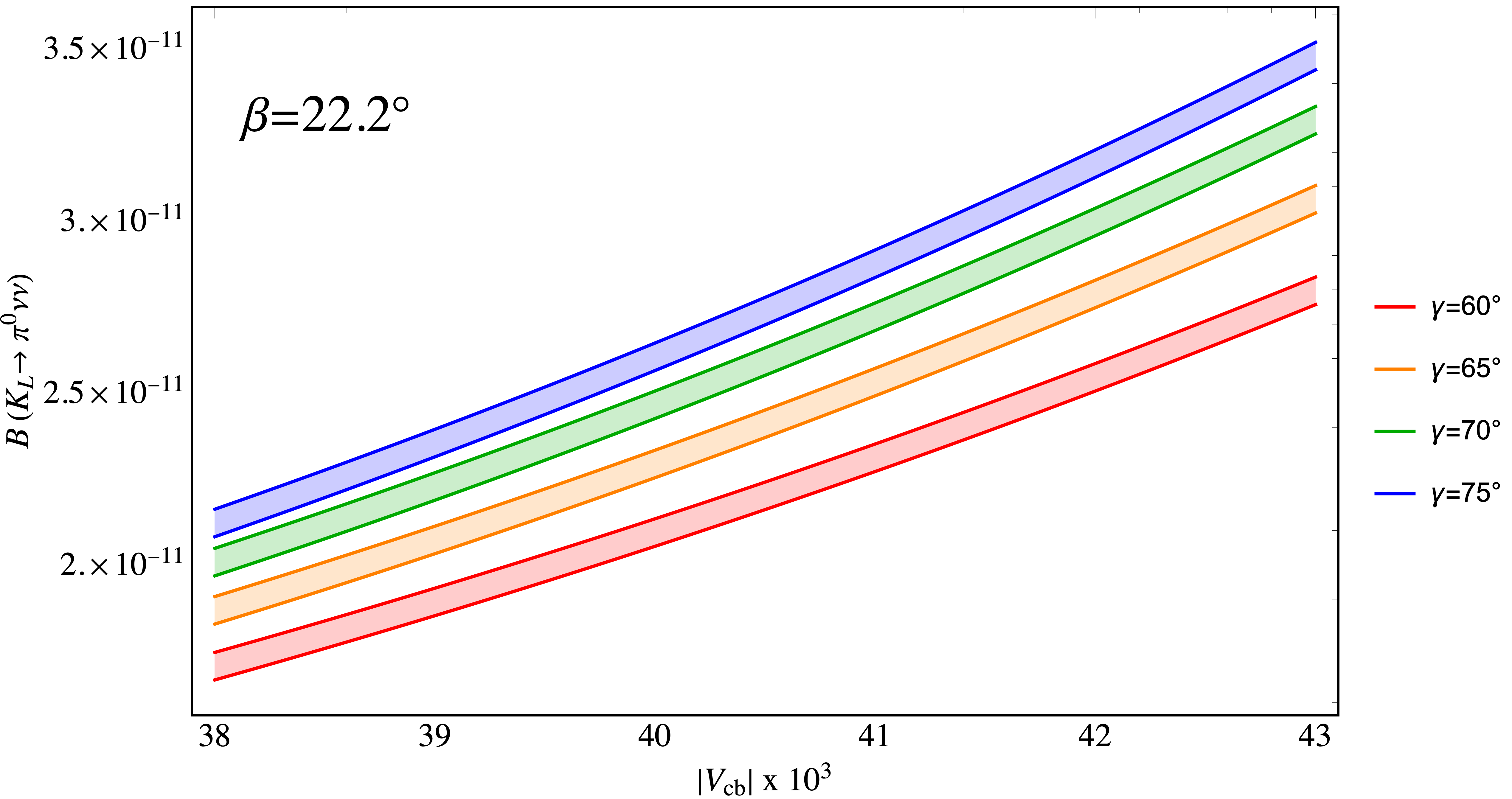}%
\caption{\it {The dependence of the branching ratios $\mathcal{B}(\kpn)$ (left panels) and $\mathcal{B}(\klpn)$ (right panels) on  $|V_{cb}|$ for different values of   $\beta=20.0^\circ,21.0^\circ, 22.0^\circ, 23.0^\circ, 24.0^\circ$  at fixed $\gamma=67^\circ$ and for different values of $\gamma=60.0^\circ, 65.0^\circ, 70.0^\circ, 75^\circ$  at fixed $\beta=22.2^\circ$ . {The width of the bands represents the uncertainties whose origin is not related to the $\gamma$, $\beta$ and $\vcb$ parameters. From \cite{Buras:2021nns}. }} 
\label{fig:CKMdependence}}
\end{figure}

In order to motivate this strategy further, I show in Fig.~\ref{fig:CKMdependence} the dependence of $\kpn$ and $\klpn$ branching ratios on $\vcb$, $\beta$ and
$\gamma$ with $\beta$ and $\gamma$ being the angles of the Unitarity Triangle
shown in Fig.~\ref{UUTa}. This dependence has been already studied with Monika Blanke
in \cite{Blanke:2018cya} but this particular figure is from the first paper
with Elena Venturini with few  details described soon  \cite{Buras:2021nns}.

One observes that the largest uncertainties on these branching ratios {are}  due to the $\vcb$ parameter: the {elimination} of this source of error is the main focus of \cite{Buras:2021nns}. This idea was already born in 2003 
in the context of $B_{s,d}\to \mu\bar\mu$ and $\Delta M_{s,d}$
\cite{Buras:2003td}. Just considering the ratio of these two observables
allowed to eliminate the dependence on the CKM parameters, in particular $\vcb$
and the weak decay constants $F_{B_{q}}$, poorly known in 2003, leaving the main
uncertainty in 
the non-perturbative parameters $\hat B_{q}$ entering $\Delta M_q$.
In 2026
these parameters are known from LQCD with respectable precision. We show some of them in Table~\ref{tab:parameters}.

\begin{table}
\centering
\begin{tabular}{|c|c|}
\hline
\hfill $m_{B_s} = 5366.8(2) \mev$  \cite{Zyla:2020zbs}   &  $m_{B_d}=5279.58(17)\mev$\hfill\cite{Zyla:2020zbs}\\
$\Delta M_s = 17.749(20) \,\text{ps}^{-1}$\hfill \cite{Zyla:2020zbs}    &  $\Delta M_d = 0.5065(19) \,\text{ps}^{-1}$\hfill \cite{Zyla:2020zbs}\\
$\beta=22.62(45)^\circ$\hfill\cite{HeavyFlavorAveragingGroupHFLAV:2024ctg}
                &  {$F_K=155.7(3)\mev$\hfill  \cite{Aoki:2019cca}}\\
        $\hat B_K=0.7627(60)$\hfill\cite{Gorbahn:2024qpe} &
 $|\eps_K|= 2.228(11)\cdot 10^{-3}$\hfill\cite{Zyla:2020zbs}\\
$F_{B_s}$ = $230.3(1.3)\mev$ \hfill \cite{Aoki:2021kgd} & $F_{B_d}$ = $190.0(1.3)\mev$ \hfill \cite{Aoki:2021kgd}  \\
$F_{B_s} \sqrt{\hat B_s}=256.1(5.7) \mev$\hfill  \cite{Dowdall:2019bea}&
$F_{B_d} \sqrt{\hat B_d}=210.6(5.5) \mev$\hfill  \cite{Dowdall:2019bea}
\\
 $\hat B_s=1.232(53)$\hfill\cite{Dowdall:2019bea}        &
 $\hat B_d=1.222(61)$ \hfill\cite{Dowdall:2019bea}
\\
$\tau_{B_s}= 1.515(4)\,\text{ps}$\hfill\cite{Amhis:2016xyh} & $\tau_{B_d}= 1.519(4)\,\text{ps}$\hfill\cite{Amhis:2016xyh}
\\
\hline
\end{tabular}
\caption {\textit{Selected input values used in $\Delta F=2$ and other flavour changing processes. For future 
    updates see FLAG \cite{Aoki:2021kgd}, PDG \cite{Zyla:2020zbs} and HFLAV
\cite{HeavyFlavorAveragingGroupHFLAV:2024ctg}.
}}
\label{tab:parameters}
\end{table}

In the spring of 2021 in a paper with Christoph Bobeth, an update of my 2003 paper, the search for NP with the help of this ratio has been reemphasized \cite{Bobeth:2021cxm}.
Finally using the 2+1+1 flavour results of \cite{Dowdall:2019bea} Elena Venturini and I
found \cite{Buras:2022wpw}
\be\label{F1}
  \boxed{R_{s\mu}=\frac{\overline{\mathcal{B}}(B_s\to\mu^+\mu^-)}{\Delta M_s}=
    (2.130^{+0.083}_{-0.053})\times 10^{-10}\text{ps}\,,}
\ee
   \be\label{F2}
   \boxed{R_{d\mu}=\frac{\mathcal{B}(B_d\to\mu^+\mu^-)}{\Delta M_d}=
(2.005^{+0.089}_{-0.066})\times 10^{-10}\text{ps}\,.}
\ee
Consequently in the SM the predictions for $B_{q}\to\mu\bar\mu$
branching ratios are rather precise as  $\Delta M_q$ have already been precisely measured. Their values together with present experimental values are given in Table~\ref{tab:SMBRs}.

In the fall of 2021 my strategy of 2003 has been generalized in collaboration with Elena to semi-leptonic
$B$ and $K$ decays \cite{Buras:2021nns}. We have constructed a number
of $\vcb$-independent ratios. Let me just list the most interesting ones.
Other can be found in the latter paper as well as in \cite{Buras:2022wpw,Buras:2022qip,Buras:2024per} and \cite{Buras:2026vbp}.

In particular one has not including
  NNLO corrections to $X(x_t)$
\be\label{F3}
  \boxed{\frac{\mathcal{B}(\kpn)}{|\varepsilon_K|^{0.82}}=(1.27\pm0.06)\times 10^{-8}{\left(\frac{\sin\gamma}{\sin 64.6^\circ}\right)^{0.015}\left(\frac{\sin 22.6^\circ}{\sin \beta}\right)^{0.71},  }            }
  \ee
  \be\label{R12a}
\boxed{\frac{\mathcal{B}(\klpn)}{|\varepsilon_K|^{1.18}}=(4.03\pm 0.21)\times 10^{-8}
    {\left(\frac{\sin\gamma}{\sin 64.6^\circ}\right)^{0.03}\left(\frac{\sin\beta}{\sin 22.6^\circ}\right)^{0.9{8}}.}}
\ee

Moreover, one finds
\be\label{R1}
\boxed{\frac{\mathcal{B}(\kpn)}{\left[{\overline{\mathcal{B}}}(B_s\to\mu^+\mu^-)\right]^{1.4}}= 53.46\pm2.75\,,}
\ee

\be\label{R5}
\boxed{\frac{\mathcal{B}(\kpn)}{\left[\mathcal{B}(B^+\to K^+\nu\bar\nu)\right]^{1.4}}={(2.28\pm0.13)\times 10^{-3}\,.}}
\ee
\be\label{R7}
\boxed{\frac{\mathcal{B}(B^+\to K^+\nu\bar\nu)}{{\overline{\mathcal{B}}}(B_s\to\mu^+\mu^-)}={(1.32\pm0.07)\times 10^{3}\,.}}
\ee

As within the SM the values of $\Delta M_d$, $\Delta M_s$ and $\varepsilon_K$
are just equal to their well measured experimental values it is possible
to determine several of the branching ratios with best precision to date.
We list them in Table~\ref{tab:SMBRs}. Moreover, it is possible to
determine all CKM parameters:
\be\label{CKMBV}
\boxed{\vcb=42.5(5)\times 10^{-3}, \quad 
  \gamma=64.6(16)^\circ, \quad \beta=22.62(45)^\circ, \quad \vub=3.76(11)\times 10^{-3}\,}
\ee
and consequently
\be\label{CKMoutput2}
\boxed{\vts=41.8(4)\times 10^{-3}, \qquad \vtd=8.64(14)\times 10^{-3}\,,\qquad
{\IM}\lambda_t=1.45(5)\times 10^{-4}\,,}
\ee
\be\label{CKMoutput3}
\boxed{\bar\varrho=0.165(12),\qquad \bar\eta=0.348(11)\,,}
\ee
where $\lambda_t=V_{ts}^*V_{td}$.

Let me next stress that presently the optimal set of CKM variables is the following one \cite{Blanke:2018cya}
\be
\vus,\qquad \vcb, \qquad \beta, \qquad \gamma\,,
\ee
which avoids the use of $\vub$ as a fundamental parameter.
In this context let me 
present two simple formulae that allow, having
$(\beta,\gamma)$, to calculate the appex of the UT in no
time, but 
to my knowledge they have been presented  only recently for the first time
\cite{Buras:2023ric,Buras:2023qaf}.  
They read
\be\label{AJB23}
\boxed{\bar\varrho=\frac{\sin\beta\cos\gamma}{\sin(\beta+\gamma)},\qquad
  \bar\eta=\frac{\sin\beta\sin\gamma}{\sin(\beta+\gamma)}.}
\ee
Evidently they can be derived by high-school students, but the UT is unknown to them and somehow no flavour physicist got the idea to present them in print so far. Numerical tables for $\bar\varrho$ and $\bar\eta$ for different values of
$\beta$ and $\gamma$ are given in
\cite{Buras:2023ric,Buras:2023qaf}.

Next, taking the present experimental result for $B_s\to \mu^+\mu^-$ allows us
to determine the present $\vcb$-independent experimental values for 
the ratios in (\ref{R1}) and (\ref{R7})
that read 
\be\label{Q1}
\boxed{\frac{\mathcal{B}(\kpn)_{\rm{exp}}}{\left[{\overline{\mathcal{B}}}(B_s\to\mu^+\mu^-)\right]^{1.4}_{\rm{exp}}}=67.52\pm15.09,\qquad    \text{SM}: 53.46\pm2.75}
\ee
\be\label{Q2}
\boxed{\frac{\mathcal{B}(B^+\to K^+\nu\bar\nu)_{\rm{exp}}}{{\overline{\mathcal{B}}}(B_s\to\mu^+\mu^-)_{\rm{exp}}}=(3.8\pm1.2)\cdot 10^{3}, \qquad \text{SM}:(1.32\pm0.07)\cdot 10^{3}}
\ee
The first ratio is consistent with the SM but the second one differs significantly from the very precise SM value. Let us hope that the experimental errors on these ratios will decrease in the coming years and
the data will depart more from the SM predictions. I wish Belle II, NA62, LHCb, CMS and ATLAS experimentalists luck in measuring precisely the
involved branching ratios.

As the results presented in Table~\ref{tab:SMBRs} and the ones for CKM parameters above assume no NP in $\Delta F=2$ observables, 
a useful rapid test of this assumption  is provided by the following, practically CKM free,
SM relation between the four $\Delta F=2$ observables \cite{Buras:2022qip} 
 \be\label{RNEW}
\boxed{\frac{ |\varepsilon_K|^{1.18}}{\Delta M_d\,\Delta M_s}=(8.37\pm 0.18)\times 10^{-5}\, \left(\frac{\sin \beta}{\sin 22.62^\circ}\right)^{1.027} K\,{\rm ps^2},}
\ee
where
\be
K=\left(\frac{\hat B_K}{0.7625}\right)^{1.18} 
\left[\frac{210.6\mev}{\sqrt{\hat B_{B_d}}F_{B_d}}\right]^2\,
\left[\frac{256.1\mev}{\sqrt{\hat B_{B_s}}F_{B_s}}\right]^2\,=\, 1.00\pm0.07\,.
\ee
The dependence on $\vcb$
drops out and the one on $\gamma$ being  negligible is included in the
uncertainty varying $\gamma$ in the range $60^\circ\le\gamma\le 70^\circ$.
Inserting the experimental values of the three $\Delta F=2$ observables on the l.h.s one finds for this ratio $(8.26\pm0.06)\times 10^{-5}$. Consequently,  with the presently known values of 
$\sqrt{\hat B_{B_d}}F_{B_d}$ and $\sqrt{\hat B_{B_s}}F_{B_s}$ from HPQCD, 
$\hat B_K$ from RBC-UKQCD 
and the present value of $\beta$ from $S_{\psi K_S}$, the SM is performing in the $\Delta F=2$ sector indeed very well. However with the $2+1$ flavours from
Fermilab Lattice and MILC Collaborations (FNAL/MILC) \cite{Bazavov:2016nty}
the central
value on the r.h.s of (\ref{RNEW}) decreases to $(6.29\pm 0.18)\times 10^{-5}$
so that the fact that this ratio  agrees with the data for present values of hadronic parameters with $2+1+1$ flavours and the experimental value of $\beta$ is remarkable. Therefore, it is very important that a second lattice collaboration
performs the calculation of  $\sqrt{\hat B_{B_d}}F_{B_d}$ and $\sqrt{\hat B_{B_s}}F_{B_s}$  with $2+1+1$ flavours.

\begin{table}
\centering
\renewcommand{\arraystretch}{1.4}
\resizebox{\columnwidth}{!}{
\begin{tabular}{|ll|l|}
\hline
Decay 
& SM Branching Ratio
& Data
\\
\hline \hline
 $B_s\to\mu^+\mu^-$ &  $(3.78^{+ 0.15}_{-0.10})\cdot 10^{-9}$      &  $(3.45\pm0.29)\cdot 10^{-9}$ \cite{HFLAV:2022pwe} 
\\
 $B_d\to\mu^+\mu^-$ &  ${(1.02^{+ 0.05}_{-0.03})}\ \cdot 10^{-10}$      & $\le 2.05\cdot 10^{-10}$ \cite{LHCb:2021awg}
\\
$B^+\to K^+\nu\bar\nu$ & $(4.99\pm 0.30)\cdot 10^{-6}$
&    $ (13\pm 4)\cdot 10^{-6}$ \cite{Belle-II:2023esi}
\\
$B^0\to K^{0*}\nu\bar\nu$ & ${(10.25\pm 0.92)}\cdot 10^{-6}$ &
 $\le 1.5\cdot 10^{-5}$ \cite{Grygier:2017tzo}
\\
\hline
$\kpn$ & $(8.65\pm 0.42)\cdot 10^{-11}$ &  $(9.6^{+1.9}_{- 1.8})\cdot 10^{-11}$  (NA62)
\\
 $\klpn$ & $(3.05\pm 0.17)\cdot 10^{-11}$ &   $\le 2.0\cdot 10^{-9}$ \cite{Ahn:2018mvc} 
\\
$(\ksm)_{\rm SD}$& {$(1.85\pm 0.12)\cdot 10^{-13}$} &   $\le 2.1\cdot 10^{-10}$
\cite{LHCb:2020ycd}
\\
\hline
\end{tabular}
}
\renewcommand{\arraystretch}{1.0}
\caption{\label{tab:SMBRs}
  \small
  Present most accurate  SM estimates  of the branching ratios  obtained using the BV-strategy. These are {\it the Magnificant Seven} decays, that being
  theoretically cleanest are optimal for the search for NP. Few numbers differ from the similar table in  \cite{Buras:2024per} due to the modification of $\beta$ and the NNLO QCD corrections to $X(x_t)$ (Talk of Emanuel Stamou at Kaon 25). 
    }
\end{table}

%\newpage
\section{DNA Strategy}\label{DNA}
The strategy just presented allows to identify quickly the ratios of
observables affected by NP. In order to find out what this NP could be
we have to do more.

In my view the fastest route to select the
candidates for NP models before doing global fits is the strategy 
developed in collaboration with Jennifer Girrbach-Noe \cite{Buras:2013ooa}. 
The main idea 
is to study first the patterns of anomalies observed in 
the data and compare them  with the patterns of deviation from  SM
predictions in a given NP scenario. Such patterns, that expose suppressions and enhancements of various observables relative to SM predictions, can be considered as DNAs of the animalcula hunted by us.
In particular the
correlations between various enhancements and suppressions can rule out
some NP scenarios already before any global fit is performed. This
strategy proposed by us  in
2013 \cite{Buras:2013ooa} has been documented in several subsequent
papers, in particular in Chapter 19.4 of my book \cite{Buras:2020xsm} and recently in \cite{Buras:2024mhy,Buras:2026vbp} with many colourful plots. Still I want to summarize its main steps and show some of these plots.

{\bf Step 1}

We construct a chart showing different observables, typically 
a branching ratio for a given decay or an asymmetry, like CP-asymmetries 
$S_{\psi K_S}$ and $S_{\psi\phi}$ and quantities $\Delta M_s$, $\Delta M_d$, 
$\varepsilon_K$ and $\epe$. The important point is to select 
the optimal set of observables which are simple enough so that definite 
predictions in a given theory can be made and can be measured in the
coming decades. The ones in Table~\ref{tab:SMBRs} are examples of such
observables among the rare K and B decays.

{\bf Step 2}

In a given NP model  we calculate the selected observables and investigate 
whether a given observable is {\em enhanced or suppressed} relative to the SM 
prediction or is 
basically unchanged. What this means requires a measure, like three 
$\sigma$.  For these three situations one can use the following colour 
coding:
\be
{\rm  \colorbox{yellow}{enhancement}} \qquad {\rm \framebox{no~change}}  \qquad {\rm 
\colorbox{black}{\textcolor{white}{\bf suppression}}}.
\ee
To this end the predictions within the SM have to be known precisely.

{\bf Step 3}

\begin{figure}[!tb]
\centering
\includegraphics[width = 0.98\textwidth]{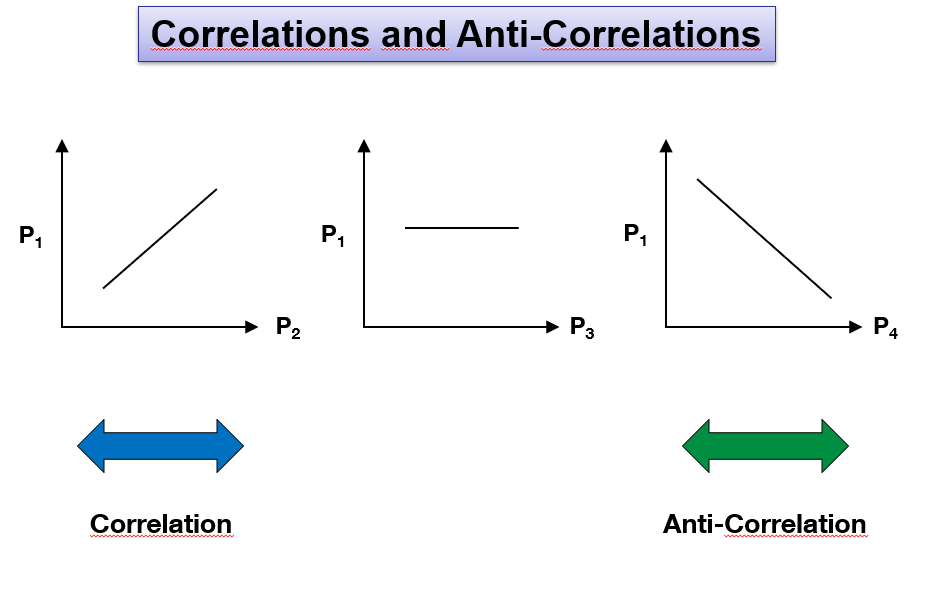}
\caption{\it Correlations, no correlations and anti-correlation. $P_i$ is
  probability representing in public colloquia the branching ratio}
 \label{Picorr}~\\[-2mm]\hrule
\end{figure}

It is only seldom that a given observable in  a given theory is uniquely 
suppressed or enhanced but frequently two observables are correlated or
uncorrelated with each other. That is the enhancement of one observable implies uniquely
an enhancement (correlation) or suppression (anti-correlation) of another 
observable. It can also happen that no change in the value of a given 
observable implies no change in another observable. This is illustrated in
Fig.~\ref{Picorr}.

The idea then is to connect in our DNA-chart 
a given pair of observables that are correlated with each other
by a line. Absence of a line means that two given observables are 
uncorrelated. In order to distinguish the correlation from anti-correlation 
one can use the following colour coding for the lines in question:
\be
%\[
{\rm correlation:} ~ \filledarrowb \qquad ~ {\rm anti - correlation:} ~ \filledarrowg . 
%\]
\ee

Let us consider a number of prominent rare decays and divide them 
into two classes:

{\bf Class A:} Decays that are governed by vector ($V=\gamma_\mu$) quark couplings. These are for instance
\be
\kpn, \qquad \klpn, \qquad B\to K\nu\bar\nu, \qquad B\to K \mu^+\mu^-.
\ee
In this case the change from left-handed to right-handed quark couplings does not 
introduce any change of the sign of NP contribution relatively to the SM one.

{\bf Class B:} Decays that are governed by axial-vector ($A=\gamma_\mu\gamma_5$) quark couplings. These are for instance
\be
K_L\to\mu^+\mu^-, \qquad B\to K^*\nu\bar\nu, \qquad B_{s,d}\to \mu^+\mu^-, \qquad B_d\to K^* \mu^+\mu^-.
\ee

In this case the change from left-handed to right-handed couplings implies
 the sign flip of NP contribution relatively to the SM one. Strictly speaking 
in the case of  $B\to K^*\nu\bar\nu$ and $B_d\to K^* \mu^+\mu^-$ this rule only  applies if the contributions from the longitudinal and parallel transversity components dominate. For perpendicular component there is no sign flip.

Thus if there is a correlation between 
two observables, one belonging to class A and the other to class B, in the presence of left-handed couplings, it is changed into anti-correlation when right-handed couplings are at work. This difference allows then 
to probe whether one deals with left-handed or right-handed couplings. Of 
course if both left-handed and right-handed couplings are involved the 
structure of correlations is modified, but still studying it one can in 
principle extract the relative size of these couplings from the data. Moreover, 
if there is a correlation or anti-correlation of two observables belonging to 
one class, the flip of sign of $\gamma_5$ will not have an impact on these 
relations, but can of course have an impact on whether a given observable is 
suppressed or enhanced relative to the SM prediction.

A graphical representation of these properties are the DNA charts \cite{Buras:2013ooa} which we will briefly discuss now.
Let me then end this description by four charts presented in
\cite{Buras:2013ooa} where further details can be found.

\begin{figure}[!tb]
\centering
\includegraphics[width = 0.49\textwidth]{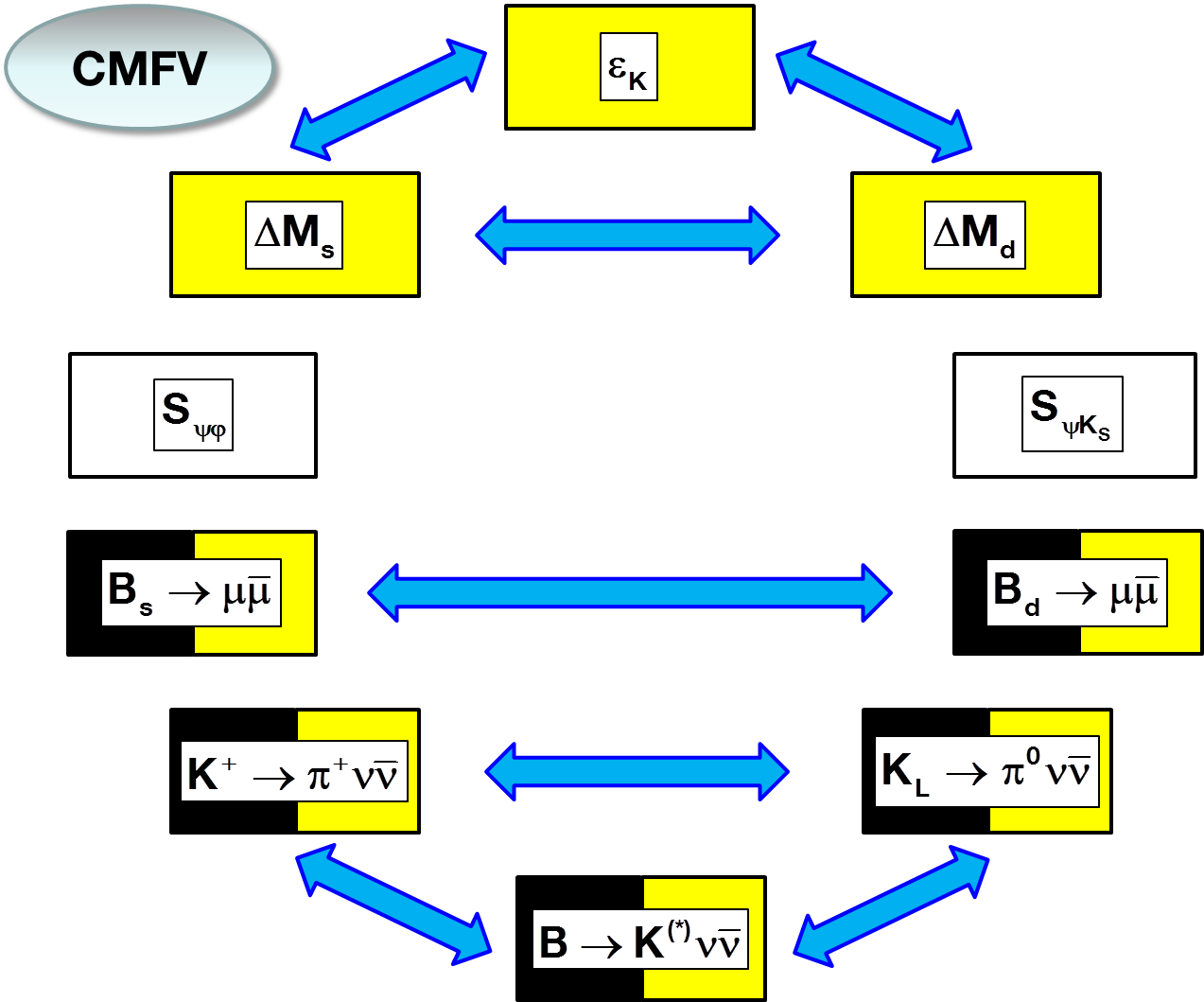}
\includegraphics[width = 0.49\textwidth]{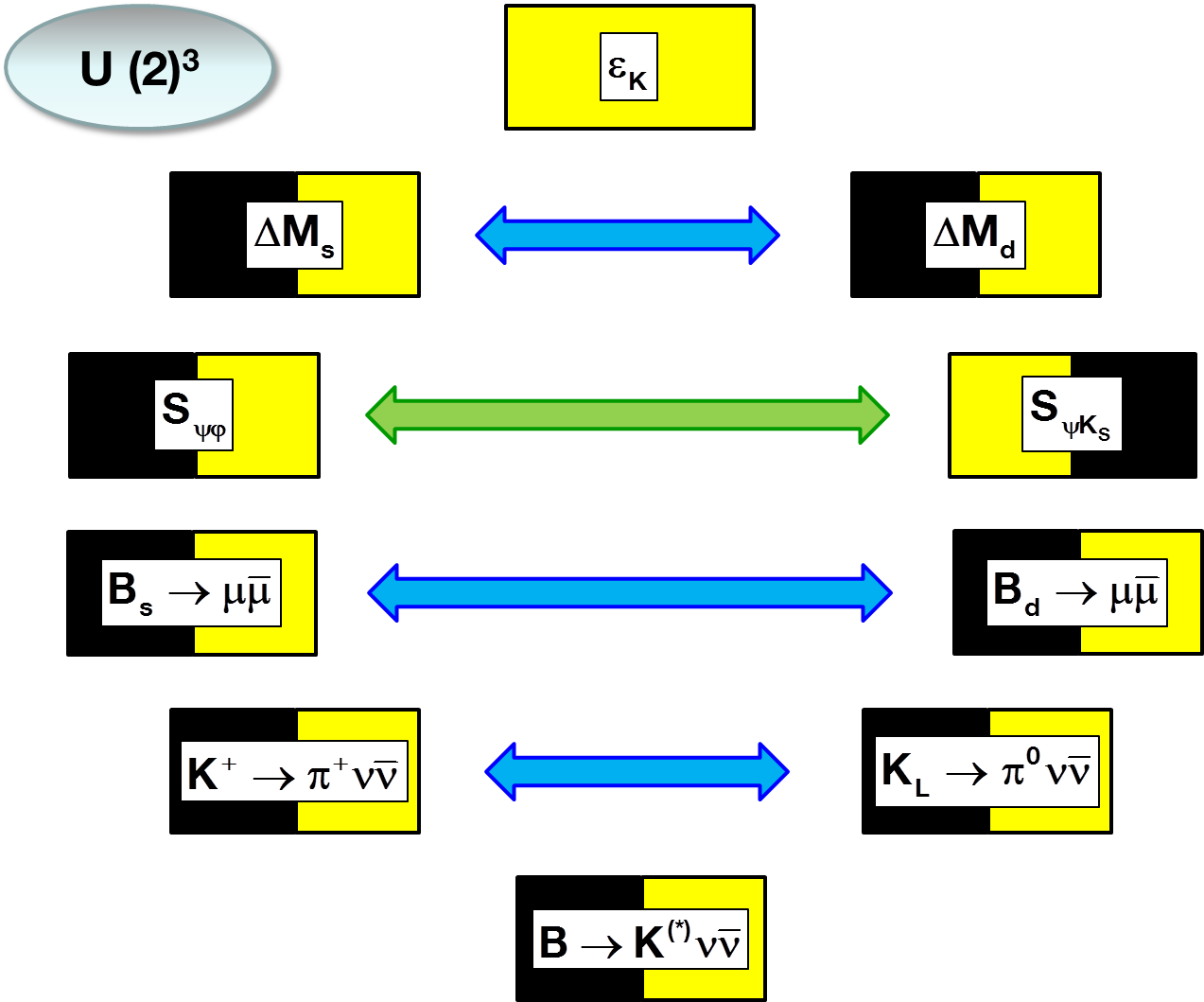}
\caption{\it DNA-chart of MFV  models (left) and of  $\text{U(2)}^3$ models (right). Yellow means   \colorbox{yellow}{enhancement}, black means
\colorbox{black}{\textcolor{white}{\bf suppression}} and white means \protect\framebox{no change}. Blue arrows
\textcolor{darkblue}{$\Leftrightarrow$}
indicate correlation and green arrows \textcolor{darkgreen}{$\Leftrightarrow$} indicate anti-correlation. From \cite{Buras:2013ooa}.}
 \label{fig:CMFVchart}~\\[-2mm]\hrule
\end{figure}

\begin{figure}[!tb]
\centering
\includegraphics[width = 0.49\textwidth]{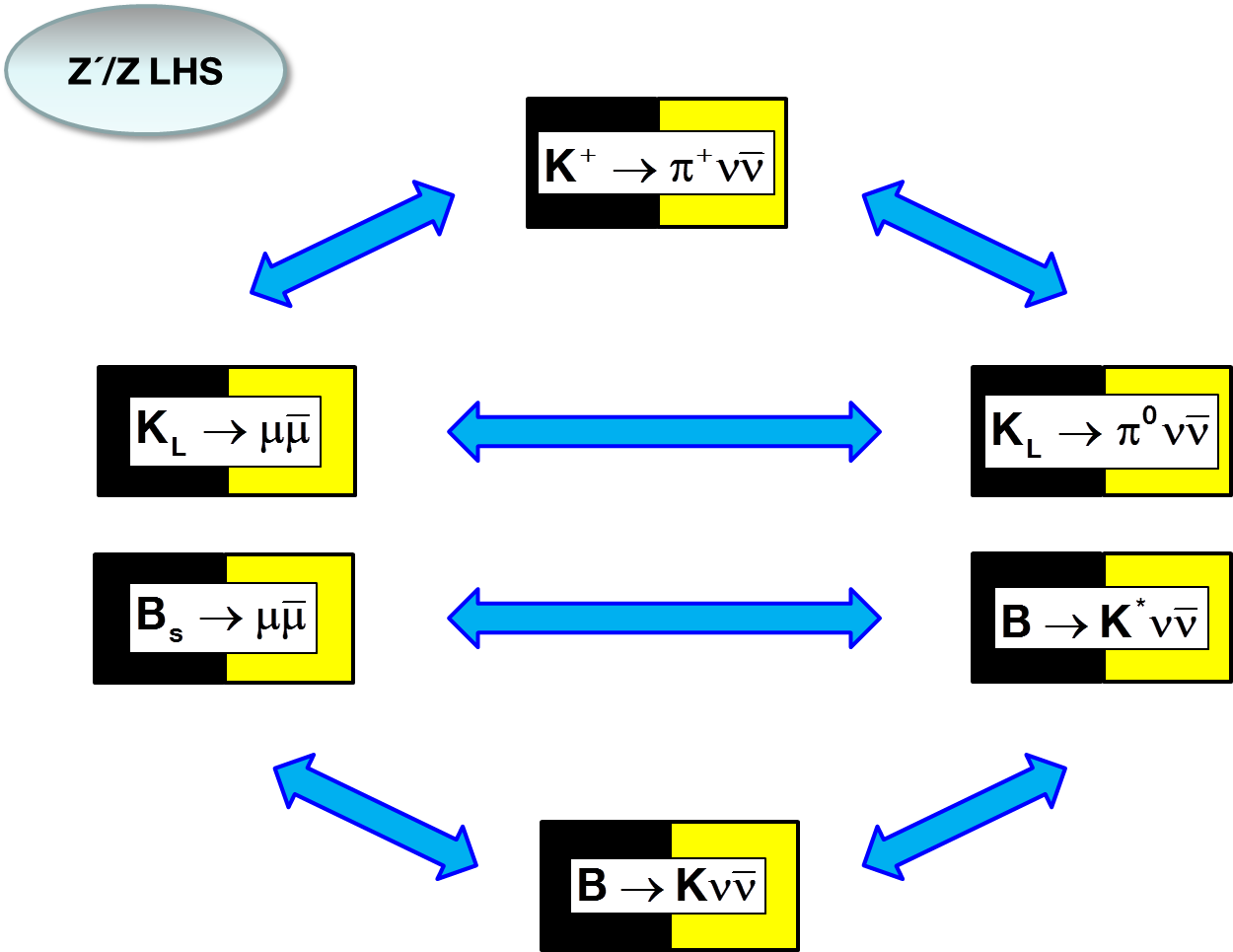}
\includegraphics[width = 0.49\textwidth]{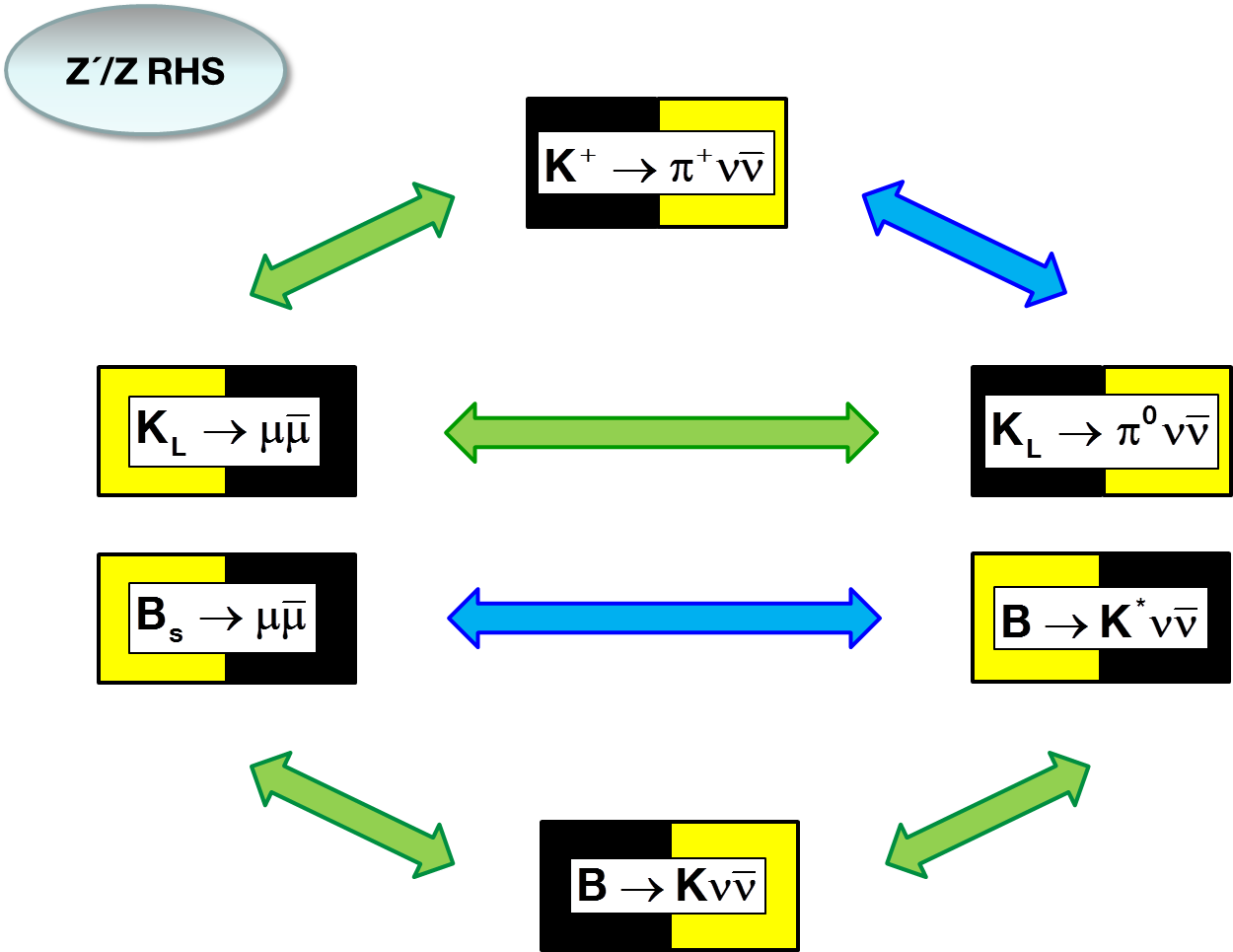}
\caption{\it DNA-charts of $Z^\prime$ models with LH and RH currents.  Yellow means   \colorbox{yellow}{enhancement}, black means
\colorbox{black}{\textcolor{white}{\bf suppression}} and white means \protect\framebox{no change}. Blue arrows
\textcolor{darkblue}{$\Leftrightarrow$}
indicate correlation and green arrows \textcolor{darkgreen}{$\Leftrightarrow$} indicate anti-correlation.  From \cite{Buras:2013ooa}.}
 \label{fig:ZPrimechart}~\\[-2mm]\hrule
\end{figure}

\begin{figure}[!tb]
  \includegraphics[width=0.49\textwidth]{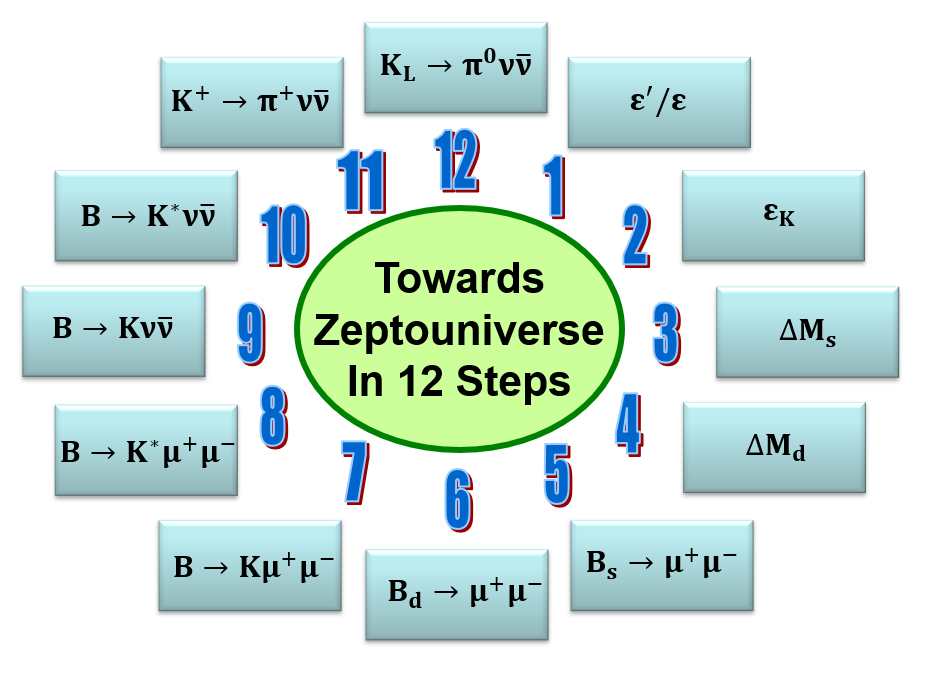}
  \includegraphics[width=0.49\textwidth]{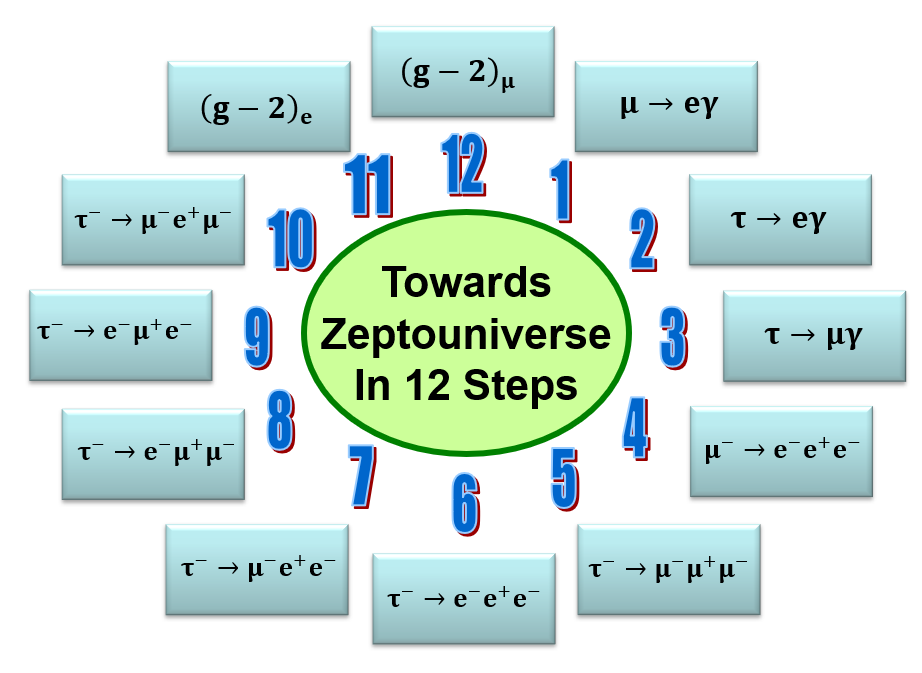}
  \includegraphics[width=0.49\textwidth]{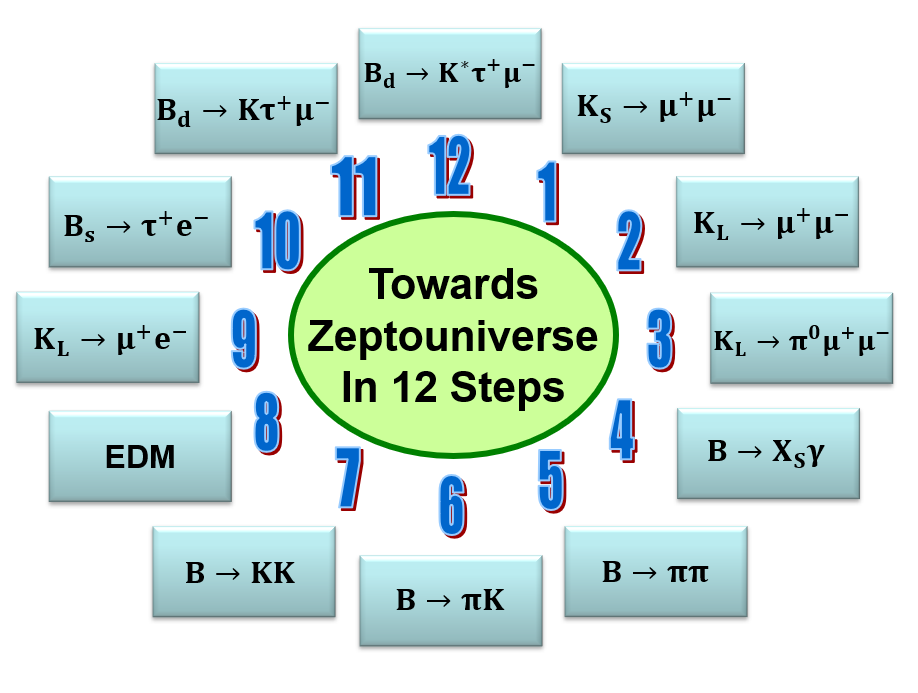}
  \includegraphics[width=0.49\textwidth]{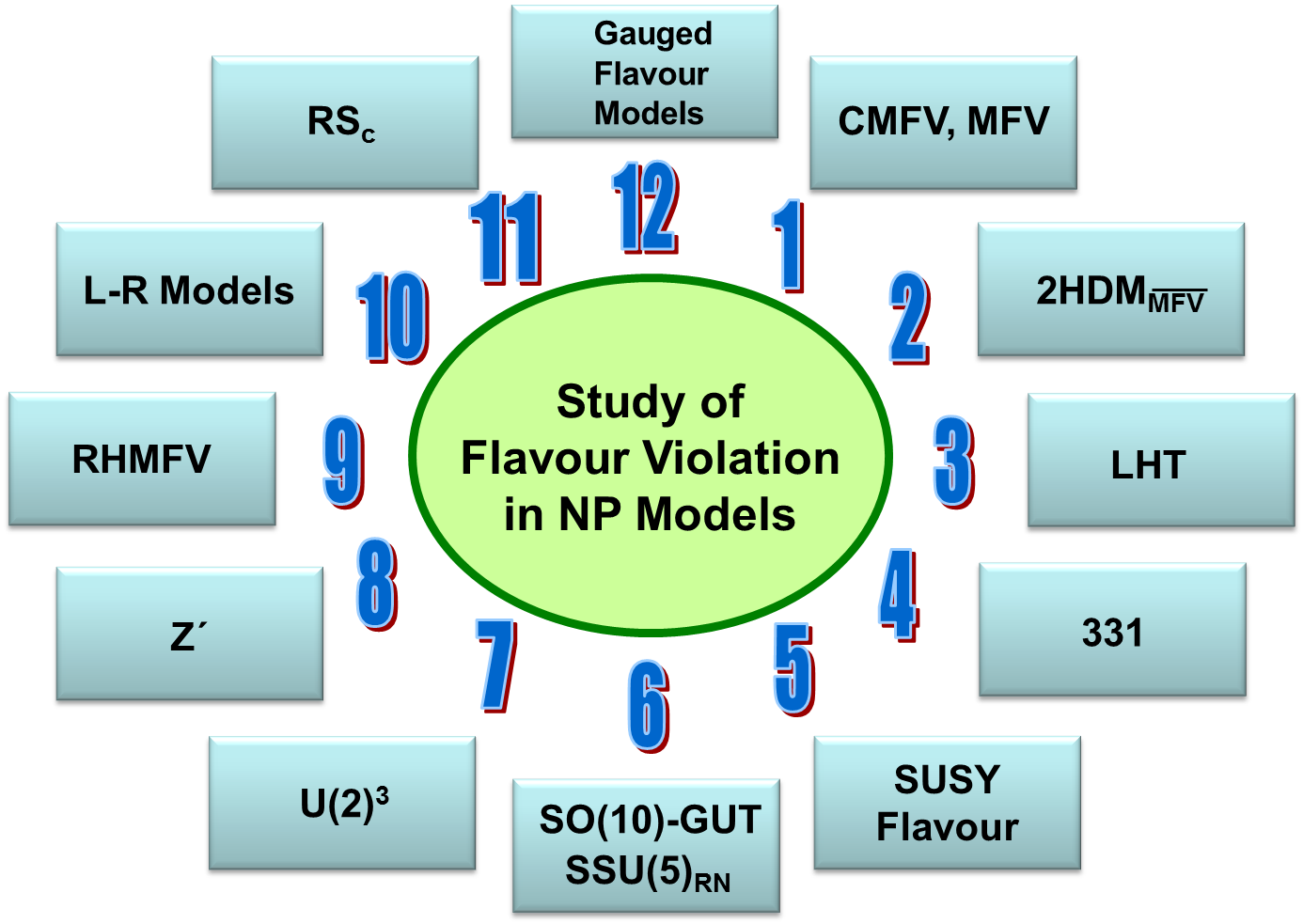}
\caption{\it Towards the Zeptouniverse in 12 Steps.}\label{Fig:1}~\\[-2mm]\hrule
\end{figure}

\begin{figure}[!tb]
  \centerline{\includegraphics[width=0.98\textwidth]{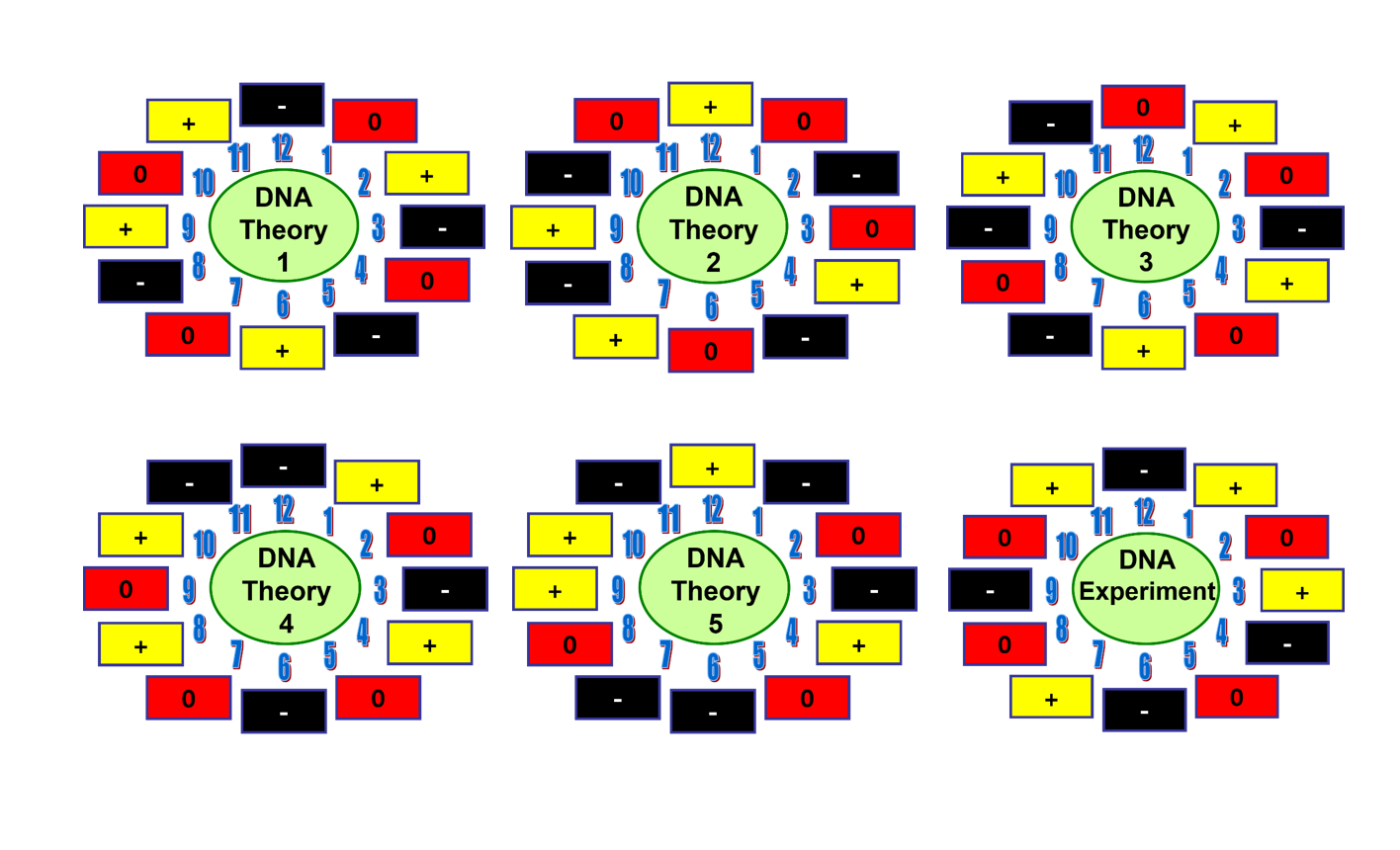}}
  \vspace{-1.0cm}
\caption{\it DNA Tests of several theories with the last one being experimental DNA.}
\label{Fig:2}~\\[-2mm]\hrule
\end{figure}

One can check that these charts summarize compactly the 
(anti-) correlations between processes of class A and B that we discussed before and also the correlations and anti-correlations within each class. In particular, the change of a correlation into an anti-correlation between two observables belonging to two different classes, when left-handed couplings 
are changed to right-handed ones, are clearly visible in these charts. We observe the following features:
\begin{itemize}
\item
Comparing   the DNA-charts of CMFV and $\text{U(2)}^3$ models 
in  Fig.~\ref{fig:CMFVchart}  we observe that the correlations 
between $K$ and $B_{s,d}$ systems are absent in the $\text{U(2)}^3$ case as the flavour symmetry is reduced from 
$\text{U(3)}^3$ down to $\text{U(2)}^3$. 
\item
As the decays $\kpn$, $\klpn$ and $B\to K\nu\bar\nu$  belonging to class A are only sensitive
to the vector quark currents, they do not change when the couplings are changed from  left-handed to right-handed ones. On the other hand, the remaining 
three decays in   Fig.~\ref{fig:ZPrimechart} belonging to class B are sensitive to axial-vector 
couplings implying interchange of enhancements and suppressions when going from 
$L$ to $R$ and also change of correlations to anti-correlations between the 
latter three and the former three decays. Note that the correlation between 
$B_s\to\mu^+\mu^-$  and $B\to K^*\mu^+\mu^-$ does not change as both decays are  sensitive only to axial-vector coupling if in the latter case the contribution 
from the longitudinal and parallel transversity components dominate.
\item
However, it should be remarked that in order to obtain the correlations or 
anti-correlations in LHS and RHS scenarios it was assumed in the DNA charts 
presented here that the signs 
of the left-handed couplings to neutrinos and the axial-vector couplings 
to muons are the same which does not have to be the case. If they are 
opposite the correlations between the decays with neutrinos and muons in 
the final state change to anti-correlations and vice versa. 
\item
On the other hand, due to $\text{SU(2)}_L$ symmetry the left-handed $Z^\prime$
 couplings to muons and neutrinos are equal and this implies the relation
\be\label{SU2}
\Delta_{L}^{\nu\bar\nu}(Z')=\frac{\Delta_V^{\mu\bar\mu}(Z')-\Delta_A^{\mu\bar\mu}(Z')}{2}. 
\ee
Therefore, once two of these couplings are determined, the third follows uniquely without the freedom mentioned in the previous item.
\item
In the context of the DNA-charts in  Fig.~\ref{fig:ZPrimechart}, the correlations involving $\klpn$ apply only if NP contributions carry some CP-phases. If this is not the case the branching ratio for $\klpn$ will remain unchanged relative to the SM one.
\end{itemize}

If in the case of tree-level $Z^\prime$ and $Z$ exchanges 
both LH and RH quark couplings are present and are equal to each 
other (LRS scenario) or differ by sign (ALRS scenario), then one finds 
\cite{Buras:2012jb}
\begin{itemize}
\item
In LRS NP contributions to $B_{s,d}\to\mu^+\mu^-$ vanish, but they are present 
in  $\klpn$, 
$\kpn$, $B_d\to K\mu^+\mu^-$ and $B\to K\nu\bar\nu$.
\item
In ALRS NP contributions to $B_{s,d}\to\mu^+\mu^-$ are non-vanishing. 
 On the other hand 
they are absent in the case of  $\klpn$, $\kpn$, $B_d\to K\mu^+\mu^-$ and 
 $B\to K\nu\bar\nu$.
\item
In  $B_d\to K^*\mu^+\mu^-$ and  $B\to K^*\nu\bar\nu$ this rule is more complicated as already stated above, but generally the LH and RH contributions interfere
destructively in LRS and constructively in ALRS. The details depend on form factors.
\end{itemize}

\begin{figure}[tb]
  \centerline{\includegraphics[width=0.98\textwidth]{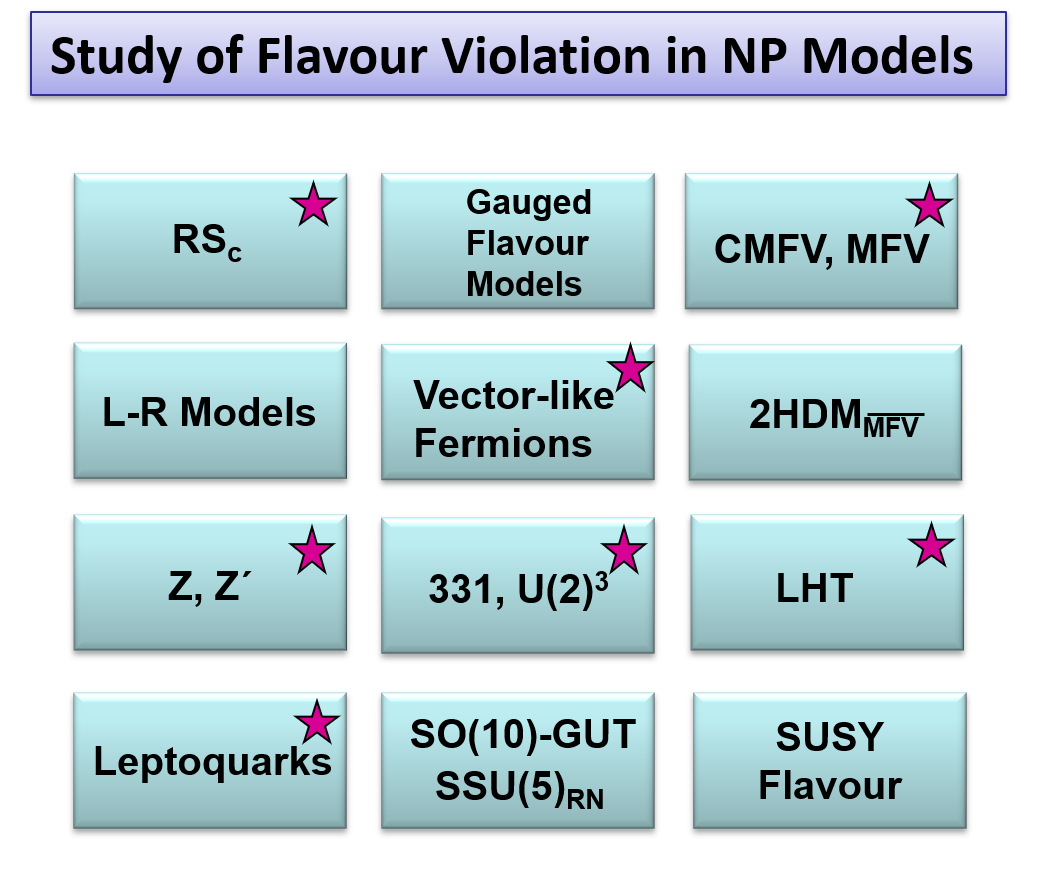}}
  \vspace{-0.5cm}
 \caption{\it NP Models investigated in my group at TUM.}
\label{Fig:NP}~\\[-2mm]\hrule
\end{figure}

As described in details in \cite{Buras:2015nta} this strategy can also be presented in a different manner.
In Fig.~\ref{Fig:1}  various observables of interest are
exposed around the first three clocks. In the last clock several NP models (theories) 
are listed for which the first three clocks can be  constructed one day.

This is ilustrated in  Fig.~\ref{Fig:2} which shows DNA tests of different
theories with enhancements, suppressions and no changes for specific observables
in a given clock of Fig.~\ref{Fig:1}. The last clock in Fig.~\ref{Fig:2}
shows the experimental result for the chosen observables. As one can see
all five theories are ruled out.

Of course because of free parameters in a given theory, it is seldom that
the signs in these clocks are unique. But as in the special NP scenarios
presented in Figs.~\ref{fig:CMFVchart} and \ref{fig:ZPrimechart} the signs
in question could be correlated. In the future when many observables
will be measured, the clocks presented here together with the DNA charts
could help to present results in a given theory in an artistic manner.

\begin{figure}[t]
\centering%
\includegraphics[width=0.6\textwidth]{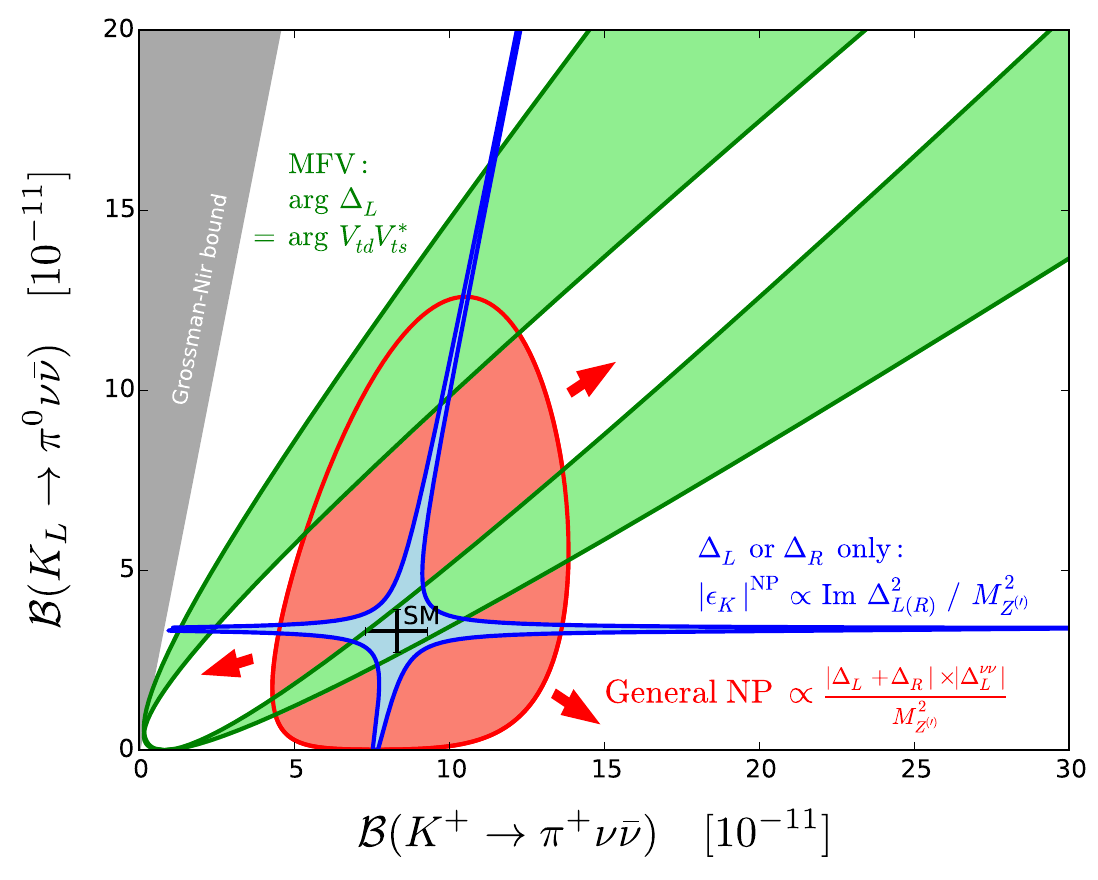}%
\caption{\it Illustrations of common correlations in the $\mathcal{B}(\kpn)$ versus $\mathcal{B}(\klpn)$ plane. See text for explanations. From \cite{Buras:2015yca} .\label{fig:illustrateEpsK}}
\end{figure}

\section{New Physics Models}\label{NPMOD}
While the strategies for searching for NP are very important, at the end
one has to construct specific models that explain possible anomalies observed
in the experimental data. Numerous NP models have been developed in the literature over the last three decades\footnote{See \cite{Albertus:2026fbe}.}.  Many of them have been analyzed in my group
at TUM, dominantly in the first two decades of this millenium. They are listed
in Fig.~\ref{Fig:NP}. The ones with stars were most popular in the literature.

My PhD students and
Postdocs contributed in a crucial manner to these analyses. We
analyzed all these NP models in details.
This means deriving Feynman Rules, calculating flavour observables and
studying phenomenological implications of these models. In particular through
correlations between different observables.

The Atlas of these analyses, with short descriptions of main results, is presented in Part X of \cite{Buras:2026vbp}
and a more extensive list can be found in chapters 15 and 16 in my book
\cite{Buras:2020xsm}. I will not repeat this list here. Instead I
will present in the next section an album of the correlations between
various observables in specific models. 

Personally, I hope that one day a $Z^\prime$, Leptoquarks and vector-like
quarks and leptons will be discovered.

\begin{figure}[tb]
\centering
\includegraphics[width=0.45\textwidth]{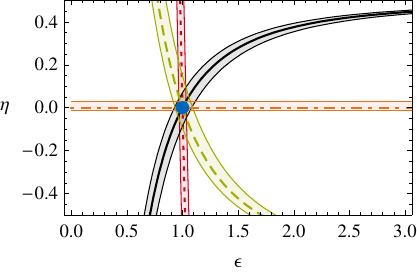}\qquad
\includegraphics[width=0.45\textwidth]{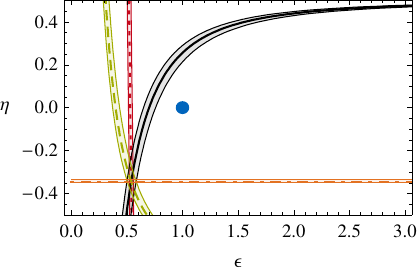}
\caption[]{\small Hypothetical constraints on the $\epsilon$-$\eta$-plane, assuming all four observables have been measured with infinite precision. See text for explanations. From \cite{Altmannshofer:2009ma}.}
\label{fig:exp-hypo}
\end{figure}

\section{Correlations in various NP Scenarios}\label{CORR}
\begin{figure}[tb]
 \centering
\includegraphics[width = 0.45\textwidth]{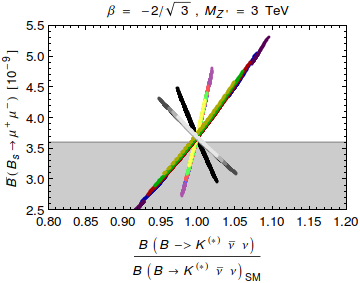}
\includegraphics[width = 0.45\textwidth]{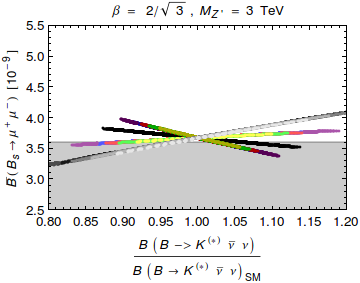}

\includegraphics[width = 0.45\textwidth]{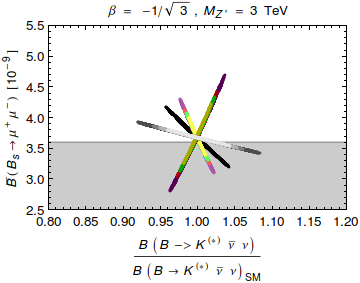}
\includegraphics[width = 0.45\textwidth]{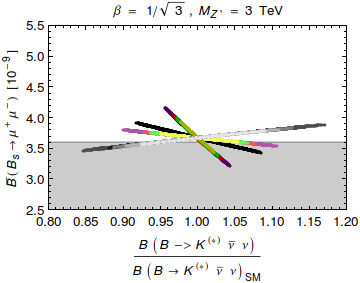}
\caption{ \it $\overline{\mathcal{B}}(B_s\to\mu^+\mu^-)$ versus the ratio ${\mathcal{B}(B\to K \nu \bar \nu)}/{ \mathcal{B}(B\to K \nu \bar 
\nu)_{\rm SM}}$  for all four $\beta=\pm\frac{2}{\sqrt{3}},
\pm
\frac{1}{\sqrt{3}}$. Colours describe different values of $\Delta M_s/(\Delta M_s)_{\rm SM}$. From \cite{Buras:2014yna}.
}\label{fig:BsmuRnuF2}~\\[-2mm]\hrule
\end{figure}

\begin{figure}[ptb]
\centering
\includegraphics[height=0.8\textheight]{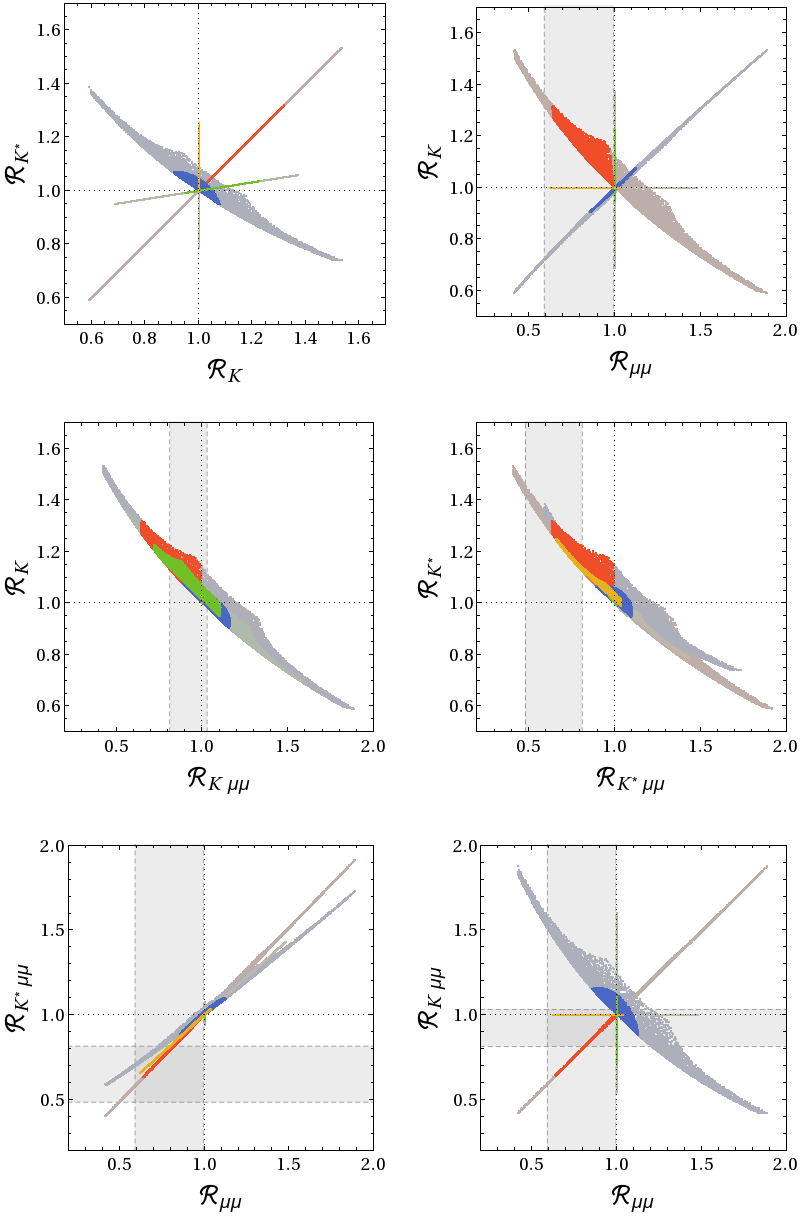}
\caption{Various correlations between observables in  LHS (red), RHS (blue), LRS (green), ALRS (yellow), assuming LFU and $\Delta_R^{\nu\nu} = \Delta_R^{\ell\ell}=0$. From \cite{Buras:2014fpa}. Update in \cite{Buras:2024mnq}.}
\label{fig:Zprime}
\end{figure}

\begin{figure}[th]
\centering
\includegraphics[width=0.75\textwidth]{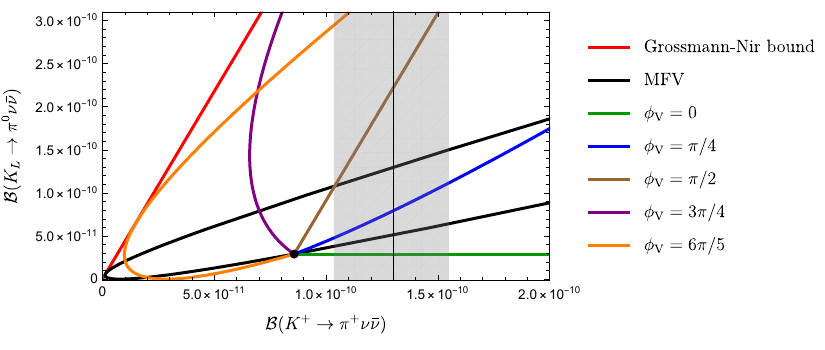}\\
\includegraphics[width=0.75\textwidth]{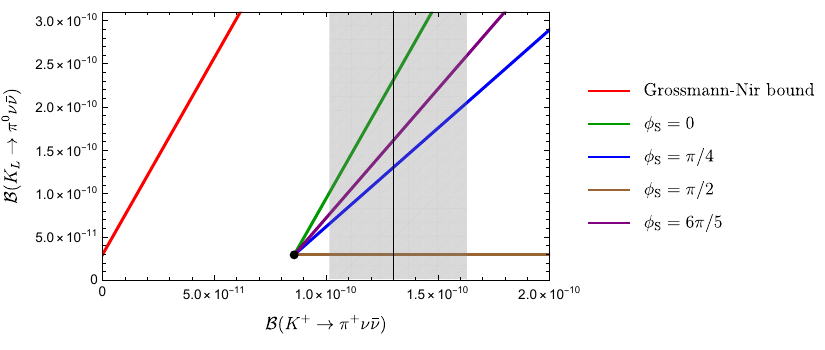}
\caption{$\mathcal{B}(\kpn)-\mathcal{B}(\klpn)$-plane for
  vector-current contributions with $C_S=0$ (top) and scalar-current contributions with $C_V=0$ (bottom) \cite{Buras:2024ewl}.}
\label{fig:KaonVectorScalar}
\end{figure}

\begin{figure}[!tb]
\centering%
\includegraphics[width = 0.45\textwidth]{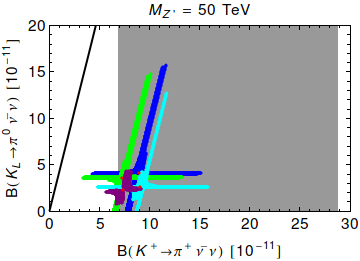}
\includegraphics[width = 0.45\textwidth]{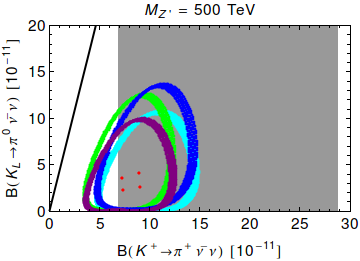}
\caption{\it $\mathcal{B}(\klpn)$ versus
$\mathcal{B}(\kpn)$ for $M_{Z^\prime} = 50~{\rm TeV}$ in the LHS (left) and for 
  $M_{Z^\prime} = 500~{\rm TeV}$ in L+R scenario.
From \cite{Buras:2014zga}.}\label{KLKP}
\end{figure}

\begin{figure}[!th]
\centering
\includegraphics[width=0.75\textwidth]{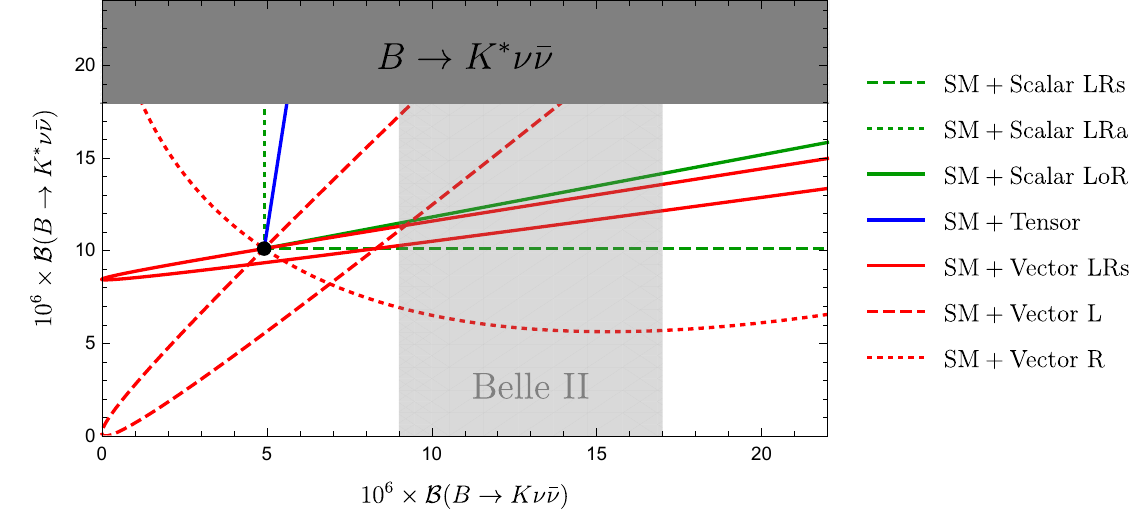}\\
\includegraphics[width=0.75\textwidth]{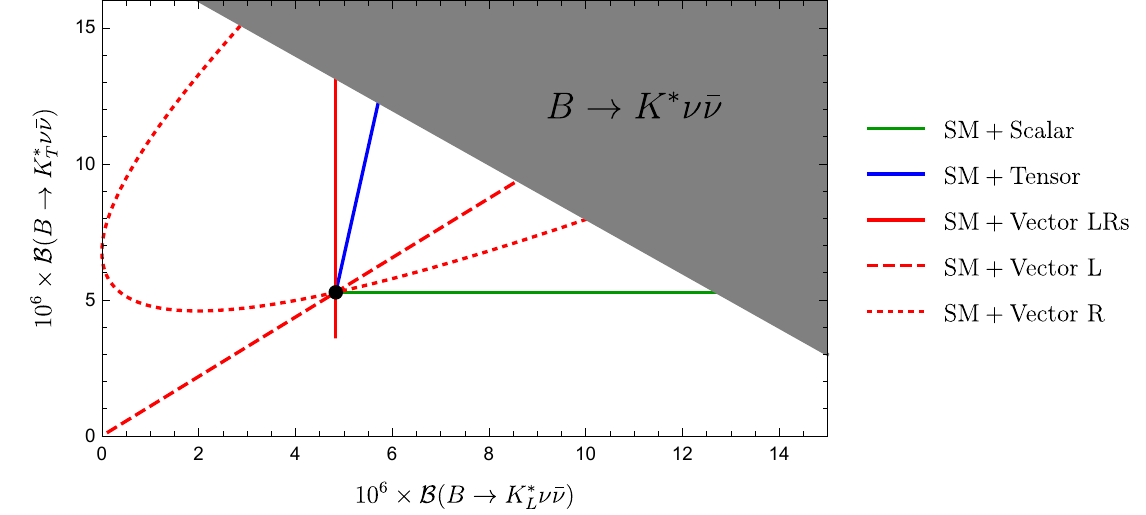}
\caption{The figure displays the $\mathcal{B}(B^+\rightarrow K^+\nu\widehat{\nu})-\mathcal{B}(B\rightarrow K^*\nu\widehat{\nu})$-plane (top) and $\mathcal{B}(B\rightarrow K^*_L\nu\widehat{\nu})-\mathcal{B}(B\rightarrow K^*_T\nu\widehat{\nu})$-plane (bottom) for different NP scenarios \cite{Buras:2024ewl}.
The SM predictions are represented by black points. The light gray region in the upper plot indicates the 2023 experimental range quoted from Belle II~\cite{Belle-II:2023esi}, 
%in (\ref{BelleIIValue}),
and the dark gray regions are excluded by the experimental limit on $B\rightarrow K^*\widehat{\nu}\nu$. The green lines show NP scenarios with scalar currents, the blue lines NP scenarios with tensor currents, and the red curves NP scenarios with vector currents.
}
\label{fig:BPlanes}
\end{figure}

In this section I show a number of correlations obtained with my PhD students,
Postdocs and other collaborators. I describe them very briefly. A bit longer
summaries can be found in my {\em Flavour Autobiography} \cite{Buras:2026vbp}.
\begin{itemize}
\item
  Fig.~\ref{fig:illustrateEpsK} illustrates  common correlations in the $\mathcal{B}(\kpn)$ versus $\mathcal{B}(\klpn)$ plane. The expanding red region illustrates the lack of correlation for models with general LH and RH NP couplings. The green region shows the correlation present in models obeying CMFV. The blue region shows the correlation induced by the constraint from $\varepsilon_K$ if only LH or RH couplings  are present. From \cite{Buras:2015yca}.
\item
  Fig.~\ref{fig:exp-hypo} presents hypothetical constraints on the $\epsilon$-$\eta$-plane \cite{Grossman:1995gt,Melikhov:1998ug}, assuming $\mathcal{B}(B \to K \nu\bar\nu)$, $\mathcal{B}(B \to K^* \nu\bar\nu)$, $\mathcal{B}(B \to X_s \nu\bar\nu)$ and $\langle F_L \rangle$ (longitudinal polarization)
  have been measured with infinite precision. $\epsilon$ and $\eta$ are
  defined through
  \begin{equation}  \label{eq:epsetadef}
 \epsilon = \frac{\sqrt{ |C^\nu_L|^2 + |C^\nu_R|^2}}{|(C^\nu_L)^\text{SM}|}~, \qquad
 \eta = \frac{-\text{Re}\left(C^\nu_L C_R^{\nu *}\right)}{|C^\nu_L|^2 + |C^\nu_R|^2}~,
\end{equation}
  where $C^\nu_{L.R}$ are the relevant WCs that enter the expressions for the four observables in question. $\epsilon=1$ in the SM. 
  A non-vanishing $\eta$ signals the presence of new right-handed down-quark flavour violating couplings which can be ideally probed by the decays in question.
  The error bands in Fig.~\ref{fig:exp-hypo} reflect the theoretical uncertainty in 2009.
The green band (dashed line) represents $\mathcal{B}(B \to K^* \nu\bar\nu)$, the black band (solid line) $\mathcal{B}(B \to K \nu\bar\nu)$, the red band (dotted line) $\mathcal{B}(B \to X_s \nu\bar\nu)$ and the orange band (dot-dashed line) $\langle F_L \rangle$.
Left: SM values for the Wilson coefficients. Right: assuming $C^\nu_L=0.5(C^\nu_L)^\text{SM}$ and $C^\nu_R=0.2(C^\nu_L)^\text{SM}$. The blue circle represents the SM point. From \cite{Altmannshofer:2009ma}.
\item
  Using the definitions in (\ref{nunuratios} and assuming that only vector currents contribute one finds \cite{Buras:2014fpa}
\begin{align}
% \label{eq:epseta-BKnn}
 \mathcal{R}_K   & = (1 - 2\,\eta)\epsilon^2
 \,, &
% \label{eq:epseta-BKsnn}
 \mathcal{R}_{K^*} 
  & =
  (1 +  \kappa_\eta \eta)\epsilon^2
  \,, &
% \label{eq:epseta-R}
 \mathcal{R}_{F_L} \equiv \frac{F_L}{F_L^\text{SM}} 
 & =  
  \frac{1+2\eta}{1+\kappa_\eta\eta}
 \,.
\label{eq:epseta-R}
\end{align}
The parameter $\kappa_\eta$ depends on the form factors. 

Since the three observables in (\ref{eq:epseta-R}) only depend on two combinations of Wilson coefficients, there is a model-independent prediction,
\begin{equation}
F_L = F_L^\text{SM}
\left(\frac{(\kappa_\eta-2)\mathcal{R}_K+4\,\mathcal{R}_{K^*}}{(\kappa_\eta+2)\mathcal{R}_{K^*}}\right)
\,.
\label{eq:FLtest}
\end{equation}
In principle, this relation can be tested experimentally (also on a bin-by-bin basis). A similar relation can be obtained for the modification of the inclusive $B\to X_s\nu\bar\nu$ branching ratio,
\begin{equation}
\text{BR}(B\to X_s\nu\bar\nu)
\approx
\text{BR}(B\to X_s\nu\bar\nu)_\text{SM}\left(
\frac{\kappa_\eta \mathcal{R}_K+2\,\mathcal{R}_{K^*}}{\kappa_\eta+2}
\right)\,.
\label{eq:Xstest}
\end{equation}
The genaralization of all these relation to the case of lepton universality
violation and/or lepton flavour violation have been presented in Section 4.5
of \cite{Buras:2024ewl}. Once the data on these four observables improves.
the sum-rule analysis presented there should offer an important insight
into possible NP at work.
\item
  Fig.~\ref{fig:BsmuRnuF2}  displays $\overline{\mathcal{B}}(B_s\to\mu^+\mu^-)$ versus the ratio ${\mathcal{B}(B\to K \nu \bar \nu)}/{ \mathcal{B}(B\to K \nu \bar \nu)_{\rm SM}}$  for  $\beta=\pm\frac{2}{\sqrt{3}},\pm\frac{1}{\sqrt{3}}$ that distinguishes between  various 331 models. Colours describe different values of $\Delta M_s/(\Delta M_s)_{\rm SM}$. From \cite{Buras:2014yna}.
\item
Fig.~\ref{fig:Zprime} from \cite{Buras:2014fpa} displays various correlations between observables in  LHS (red), RHS (blue), LRS (green), ALRS (yellow), assuming LFU and $\Delta_R^{\nu\nu} = \Delta_R^{\ell\ell}=0$. All points satisfy  $0.9\leq \Delta M_s/(\Delta M_s)_{\text{SM}}\leq 1.1$, $-0.14\leq S_{\psi\phi}\leq 0.14$.
Grey  regions are disfavoured at $2\sigma$ by $b\to s\mu^+\mu^-$ constraints in 2014. All ratios are defined
by
\be\label{nunuratios}
\mathcal{R}_{K(K^*)}=\frac{\mathcal{B}(B\to K(K^*)\nu\bar\nu)}{\mathcal{B}(B\to K(K^*)\nu\bar\nu)_{\text{SM}}},\quad
\mathcal{R}_{K(K^*)\mu\mu}=\frac{\mathcal{B}(B\to K(K^*)\mu\bar\mu)}{\mathcal{B}(B\to K(K^*)\mu\bar\mu)_{\text{SM}}},\,
\ee
and $\mathcal{R}_{\mu\mu}$ for $B_s\to\mu\bar\mu$.  The 2024 update of these correlations has been presented in \cite{Buras:2024mnq} and is summarized in \cite{Buras:2026vbp}.
\end{itemize}
%%%%%%%%%%%%%%%%%%%%%%%
\begin{itemize}
\item
Fig.~\ref{fig:KaonVectorScalar}  shows   $\mathcal{B}(\kpn)-\mathcal{B}(\klpn)$-plane for
  vector-current contributions with $C_S=0$ (top) and scalar-current contributions with $C_V=0$ (bottom) assuming lepton-flavour universality. The phase $\phi_V$ $(\phi_S)$ is fixed to different values (see legend) and $C_V$ $(C_S)$, the size of
  vector (scalar) contributions, varied.
  The red line indicates the Grossmann-Nir bound. The SM contribution is represented by a black point. The grey region represents the present experimental $1\sigma$ range. See \cite{Buras:2024ewl} for details.
\item
Fig.~\ref{KLKP} displays $\mathcal{B}(\klpn)$ versus
$\mathcal{B}(\kpn)$ for $M_{Z^\prime} = 50~{\rm TeV}$ in the LHS (left) and for 
$M_{Z^\prime} = 500~{\rm TeV}$ in L+R scenario. The colours distinguish between different CKM input considered by us which also implies the four red points corresponding to the SM central values for these four CKM scenarios.
The black line corresponds to the Grossman-Nir bound \cite{Grossman:1997sk}. The gray region shows the
experimental range of $\mathcal{B}(\kpn))_\text{exp}=(17.3^{+11.5}_{-10.5})\times 10^{-11}$ known in 2014 \cite{Buras:2014zga}. With the improved value in Table~\ref{tab:SMBRs} this region is by a factor of $2.5$ smaller.
This paper illustrates that very high
energy scales can be resolved with the help of rare Kaon decays. It will be
interesting to repeat this analysis one day when the experimental data improve.
\item
Fig.~\ref{fig:BPlanes}  displays the $\mathcal{B}(B^+\rightarrow K^+\nu\widehat{\nu})-\mathcal{B}(B\rightarrow K^*\nu\widehat{\nu})$-plane (top) and $\mathcal{B}(B\rightarrow K^*_L\nu\widehat{\nu})-\mathcal{B}(B\rightarrow K^*_T\nu\widehat{\nu})$-plane (bottom) for different NP scenarios as described in the figure caption (see \cite{Buras:2024ewl} for details).
\end{itemize}

\begin{figure}[t]
  \begin{center}
    \includegraphics[width=0.65\textwidth]{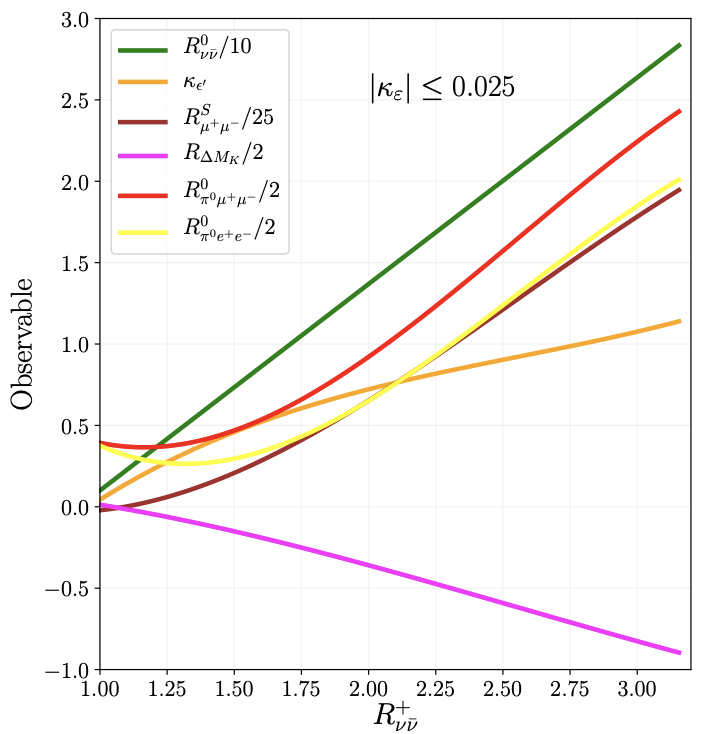}
   \caption{Correlations between the observable 
  $R_{\nu \bar \nu}^+$ and various other Kaon observables in a $Z^\prime$ model.
  All ratios $R_i=1$ in the SM. Note that for $\klpn$ and $K_S\to\mu^+\mu^-$ the
  ratios are divided by 10 and 25 respectively.  From \cite{Aebischer:2023mbz}.
  }
\label{Fig:3}
\end{center}
\end{figure}

\begin{itemize}
\item
  Fig.~\ref{Fig:3} shows the correlations between the observable 
  $R_{\nu \bar \nu}^+$ and various other Kaon observables in a $Z^\prime$ model with the imaginary $Z\bar sd$ coupling implying no NP contributions to $\varepsilon_K$.
  All ratios $R_i=1$ in the SM. Note that for $\klpn$ and $K_S\to\mu^+\mu^-$ the
  ratios are divided by 10 and 25 respectively.  From \cite{Aebischer:2023mbz}.
  \end{itemize}

In Table~\ref{tab:corr} the references to papers from my group that analyzed 
various correlations in several NP scenarios have been collected for 
convenience. The more general strategies like the ones with Jennifer \cite{Buras:2013ooa} and Elena  \cite{Buras:2021nns,Buras:2022wpw,Venturini:2022sdf}  have not been listed there. While the last entry in this table is our recent SMEFT
review \cite{Aebischer:2025qhh} in which such correlations have been
described systematically, several analyses listed in this table were performed
 in the
framework of the SMEFT. This is in particular the case of \cite{Buras:2014fpa},
\cite{Bobeth:2017xry} and \cite{Buras:2024mnq}.

\begin{table}[thb]
\begin{center}
\begin{tabular}{|c|c|}
\hline
{\bf Model} & {\bf Reference}   \\
\hline
CMFV &   \cite{Buras:2000dm,Buras:2003jf,Blanke:2006ig,Buras:2003td}\\
$\text{U(2)}^3$& \cite{Buras:2012sd,Buras:2015yca}\\
${\rm 2HDM_{\overline{MFV}}}$ &  \cite{Buras:2010mh,Buras:2010zm} \\
ACD Model & \cite{Buras:2002ej,Buras:2003mk} \\
LH        &  \cite{Buras:2004kq,Buras:2006wk}\\
LHT &  \cite{Blanke:2009am,Bigi:2009df,Blanke:2015wba}\\
SM4 &  \cite{Buras:2010pi,Buras:2010nd,Buras:2010cp}\\
AC,~RVV2,~AMK,~$\delta$LL &  \cite{Altmannshofer:2009ne,Altmannshofer:2010ad}
 \\ 
$\text{SSU(5)}_{RN}$ & \cite{Buras:2010pm}\\
FBMSSM & \cite{Altmannshofer:2008hc}\\
RHMFV & \cite{Buras:2010pz}\\
RSc & \cite{Blanke:2008yr,Blanke:2008zb}\\
Minimal Theory of Fermion Masses &\cite{Buras:2013td}\\
$Z^\prime$ &  \cite{Buras:2012jb,Buras:2014fpa,Aebischer:2023mbz,Buras:2024mnq}   \\
331 & \cite{Buras:2012dp,Buras:2014yna,Buras:2023ldz},  \\
Scalars & \cite{Buras:2013uqa,Buras:2013rqa}\\
Leptoquarks & \cite{Bobeth:2017ecx}   \\
VLQ  & \cite{Bobeth:2016llm}  \\
SMEFT & \cite{Aebischer:2025qhh} \\
\hline
\end{tabular}
\end{center}
\vspace{0.1cm}
\caption[]{References to correlations between observables in various NP models studied in my group.
\label{tab:corr}}
\end{table}


%\newpage
\section{SMEFT Correlations between Flavour Observables}\label{SMEFTCORR}
\subsection{Preface}
As we have seen in the previous section several processes are correlated within a given NP model through
various flavour and gauge symmetries present in this model. However independently of such symmetries specific to a given model,  additional correlations between
observables are generated through operator mixing in the process of RG evolution. This has often an important
impact on phenomenology as is evident from the analyses listed in Table~\ref{tab:corr} and the SMEFT review \cite{Aebischer:2025qhh}.
The importance of RG effects has been also reemphasized in \cite{Bartocci:2024fmm}.  Here we just mention two examples investigated in detail in the literature.

First, the explanation of the so-called $B$-physics anomalies in semi-leptonic
$B$ decays  with leptoquarks implies new effects in
Lepton Flavour Violating (LFV) decays
thereby putting significant bounds on the coefficients of semi-leptonic operators as stressed in particular  in \cite{Feruglio:2016gvd, Feruglio:2017rjo}.

Similarly, the attempt to obtain significant enhancement of the ratio $\epe$
within leptoquark models \cite{Bobeth:2017ecx} is very  much restricted
by  rare Kaon  decays.

Generally such correlations are within processes of a given type like
semi-leptonic,  non-leptonic and leptonic ones. But as exemplified above and
discussed at length in \cite{Aebischer:2025qhh}.
they can take place between different types of processes. 

In the following we will elaborate briefly on various origins of correlations
that are found in an effective theory at different stages of the analysis.
Much more extensive analysis with tables can be found in the  SMEFT review in
\cite{Aebischer:2025qhh}.

\subsection{Type of Correlations in the SMEFT}\label{type}

\begin{center}
{\bf\boldmath Correlations through common WCs at Tree-level}
\end{center}
First of all there are correlations when
a given WC enters two or more observables and varying its value changes the predictions of these observables in a correlated manner. Tables 17-19 in \cite{Aebischer:2025qhh}
indicate which classes of observables are correlated in this manner. These are
the strongest correlations between observables.

\begin{center}
{\bf\boldmath $\text{SU(2)}_L$ correlations at Tree-level}
\end{center}

Next there are correlations due to $\text{SU(2)}_L$ gauge symmetry.
Choosing a value of a given $C_i(\Lambda)$ implies a value of another
$C_j(\Lambda)$ that is related to $C_i(\Lambda)$ by the unbroken
$\text{SU(2)}_L$ gauge symmetry. Numerous tables in \cite{Aebischer:2025qhh}
indicate which classes of observables are correlated in this manner. In particular the decays governed by transitions $b\to s \mu^+\mu^-$ and $b\to s \nu\bar\nu$ are correlated in this manner and the same applies to $s\to d \mu^+\mu^-$ and $s\to d \nu\bar\nu$. Also such correlations are strong.

\begin{center}
  {\bf Operator Mixing due to Gauge Interactions}
\end{center}

 When RG evolution is performed from the NP scale $\Lambda_{\text{NP}}$ down to $\muEW$ and subsequently to $\muLow$
    new operators enter the game that were not active at the NP scale. They can modify the correlations present already at tree-level but could also lead to new correlations between observables that were absent at tree-level. Being suppressed by gauge couplings they
    are generally smaller than at tree-level except of course for
    those which were absent at tree level. But in the case of large anomalous dimensions (ADMs) of involved operators
    they can still have significant impact on the phenomenology through large
    logarithms multiplying the gauge couplings. Again in \cite{Aebischer:2025qhh} it is indicated  which classes of observables are correlated in this manner.

\begin{center}
  {\bf Operator and Flavour Mixing due to Top Quark Yukawas}
\end{center}

Of a particular importance is the impact of the top quark Yukawa coupling, both
on the operator mixing like in the case of gauge interactions but
in addition due to flavour mixing absent in the latter case. In particular
when starting the RG evolution at  $\Lambda_{\text{NP}}$ in a given
flavour basis in which either the up-quark Yukawa matrix $\hat Y_u$ or the down-quark Yukawa matrix $\hat Y_d$ is diagonal, one finds that at  $\muEW$ after the RG evolution, these matrices being respectively diagonal at $\Lambda_{\text{NP}}$, are no longer diagonal. A back-rotation \cite{Aebischer:2020lsx} has to be performed to make these matrices diagonal at $\muEW$. This can also have an impact not only on the values of WCs but also on  correlations between different observables.

\begin{center}
  {\bf  Matching Effects}
\end{center}

      Also the matching between a UV completion and the SMEFT as well as the
      matching between the SMEFT and the WET can have an impact on the correlations between various observables. It can be significant but in contrast to RG effects it is not enhanced by large logarithms of the type
      $\ln\Lambda_{\text{NP}}/\muEW$. This applies both to the tree-level matching and to one-loop matching. The inclusion of the latter is required in particular
      at the NLO level of a RG analysis to cancel  the unphysical
      renormalization scheme dependencies of two-loop anomalous dimensions and
      also various scale dependencies. 

\begin{center}
  {\bf Modification of SM Parametrs through dim-6 Operators}
\end{center}

      Finally, the correlations in question can enter through $1/\Lambda^2$
      effects on the SM parameters.

      This list of possible dynamical effects makes it clear that to find
      the correlations between observables within the SMEFT computer codes
      are indispensable. Fortunately, there
      exist a number of very efficient codes on the market
      \cite{Aebischer:2023nnv} that can perform
      this task. The listing of these codes with brief explanations can be found in Section 10 of \cite{Aebischer:2025qhh}.

      However, no numerical analysis of the correlations in question has been performed in \cite{Aebischer:2025qhh}. Presently two numerical analyses are in progress. There is a chance that at least one of them will be ccompleted by
      June this year.

\section{Summary, Shopping List  and Outlook}\label{OUT}
\subsection{Summary}
Despite the presence of various anomalies in the existing data, it is
clearly not evident which animalcula could be responsible for them. On the basis of the studies I have been involved in, possibly
the main candidates are $Z^\prime$ vector bosons, leptoquarks, vector-like
quarks and vector-like leptons but to find out without any doubts what they are  we need more data,
in particular for theoretically clean decays as the ones listed in Table~\ref{tab:SMBRs}. Precise measurements of branching ratios for these decays and of
various kinematical distributions should allow us in this decade and next decade to find
out which animalcula are responsible for them. The strategies presented in
this lecture and also discussed in more detail in \cite{Buras:2026vbp}
should be helpful in this respect. It should be stressed that  the ratios
listed in Section~\ref{BV}
test the SM
independenty of whether NP effects are strongly suppressed in  $\Delta F=2$ processes or not. Having several measurements of these ratios and
seeing some pattern in deviations from their SM predictions would
already be very exciting. One could then return to the strategies like the
DNA one of Section~\ref{DNA} and other strategies involving correlations
between observables. In particular updated studies of 
explicit models listed in Fig.~\ref{Fig:NP} could be very helpful in identifying
the candidates for a UV completion.

\begin{figure}[t!]
  \centering%
  \includegraphics[width=0.56\textwidth]{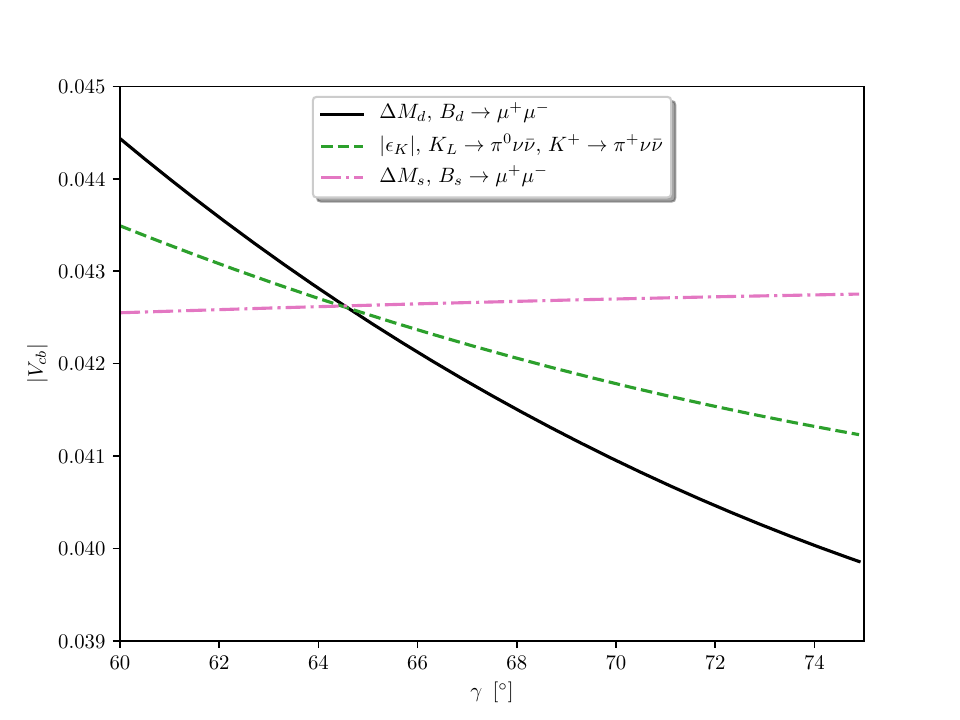}
  \includegraphics[width=0.42\textwidth]{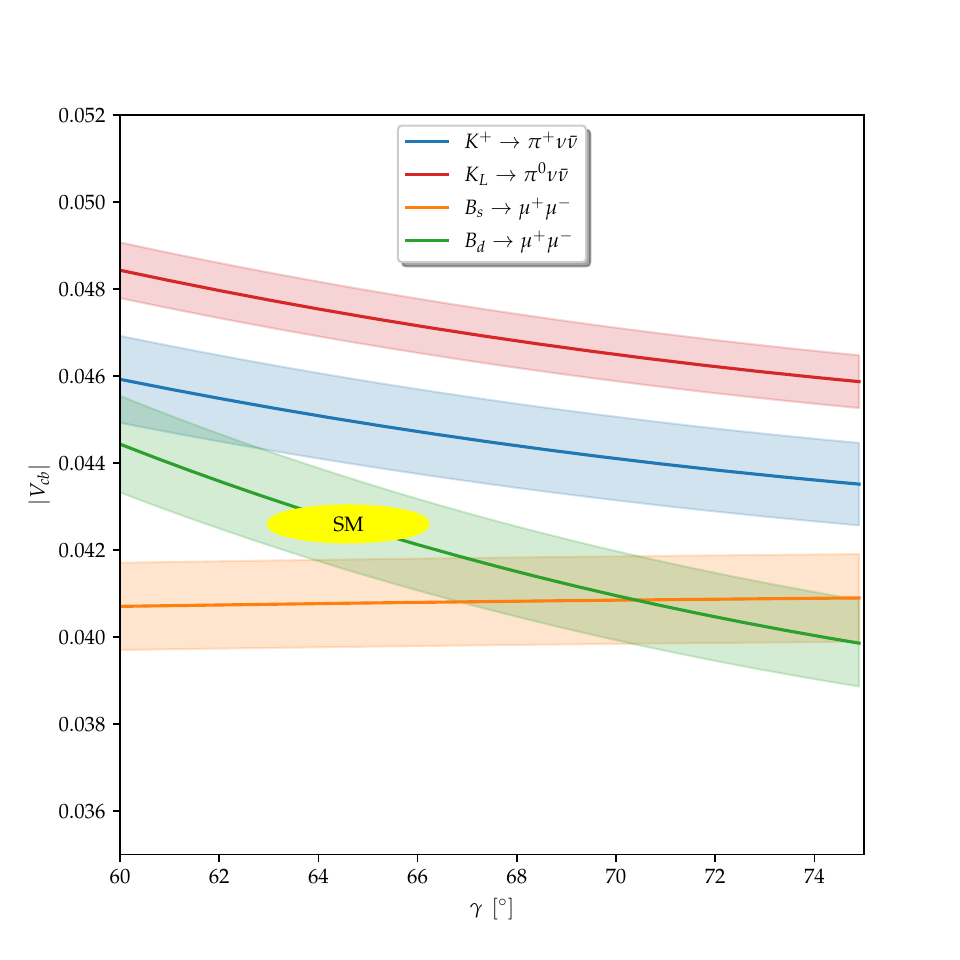}
  \caption{\it {Left:Schematic illustration of the action of the seven observables in the $\vcb-\gamma$ plane in the context of the SM. We set $\beta=22.2^\circ$ and all uncertainties to zero. Right:The impact of hypothetical future measurements of the branching ratios for $\kpn$, $\klpn$,  $B_d\to\mu^+\mu^-$ and $B_s\to\mu^+\mu^-$  on the $\vcb-\gamma$ plane. All uncertainties  are included. The yellow disc  represents the SM.                  From \cite{Buras:2022nfn}.}
\label{fig:XX}}
\end{figure}

\subsection{Shopping List}
Let me next present my shopping list for the coming years. A much more detailed list can be found in Chapter 20 of my book \cite{Buras:2020xsm}.
I list here only the entries related to indirect searches for NP, in particular
flavour physics. There is no question about that discovering a new particle
at the LHC would be most exciting as this would reduce the number of possible
extensions of the SM by much. Here comes my shortened shopping list.
\begin{enumerate}
\item
  Precise measurements of the branching ratios for the 7 magnificant decays listed in Table~\ref{tab:SMBRs}  and as studied in \cite{Buras:2024ewl}
  of missing energy distributions for four decays among them.  
Presently on a forefront are  the $B\to K^{(*)}\nu\bar\nu$ decays studied intensively by the Belle~II experiment~\cite{Belle-II:2023esi}, giving some
hints for NP contributions. There are very many recent analyses  of these data.
A collection
of references to these analyses can be found in \cite{Aebischer:2025qhh} but surely some have been performed since then.
The same applies to $\kpn$ for which a very interesting result has 
been provided by the NA62 experiment listed in Table~\ref{tab:SMBRs}.
A new one should be available this or next  year. I am looking forward
to the improved determination of
the ratios in (\ref{R1}) and (\ref{R7}). As seen in (\ref{Q1}) the first one
is consistent with the SM but could still differ from it when a more accurate
measurement from NA62 will be known. On the other hand as seen in 
(\ref{Q2}) the ratio (\ref{R7})
 differs significantly from the  precise SM prediction.
\item
  Precise measurements of the branching ratios and of other observables in $B\to K\ell^+\ell^-$ and $B\to K^*\ell^+\ell^-$ decays \cite{Altmannshofer:2008dz}
  and the clarification of the
  anomalies in them.
\item
  Precise measurements of $\vcb$ and $\gamma$ in tree-level decays that
  would provide additional tests of the BV-strategy of Section~\ref{BV}. In this context the
  $(\vcb,\gamma)$ plots in Fig.~\ref{fig:XX} could help to identify the presence of NP.
 \item
   Clarification of anomalies in $B\to\pi K$ decays \cite{Buras:2003dj,Buras:2004ub,Fleischer:2018bld,Berthiaume:2023kmp,Datta:2024zrl,Szabelski:2024cem} and
   continuation of the intensive studies of several other
   $B\to PP$ decays with $B=B_{d,s}^0,B^+$ and $P=\pi,K,K^*$.
 \item
   Completion of the calculations of $\epe$ and of the $\Delta I=1/2$ rule
   by the RBC-UKQCD lattice QCD collaboration and of corresponding calculations by another
   Lattice QCD group with $2+1+1$ flavours. In particular the isospin breaking contributions to QCD penguins are still missing in present lattice results for $\epe$.  A stressed in \cite{Buras:2023qaf},
   Jean-Marc G{\'e}rard and I expect on the basis of DQCD, developed
   with Bill Bardeen, significant NP contributions to $\epe$. 
 \item
   Calculation of hadronic matrix elements relevant for $\Delta M_s$ and
   $\Delta M_d$ with  $2+1+1$ flavours by other lattice QCD collaborations in order to check HPQCD results \cite{Dowdall:2019bea}
   for these matrix elements.
 \item
   Further searches for lepton flavour violation and EDMs.
\end{enumerate}

\begin{figure}[t]
\centering%
\includegraphics[width=0.8\textwidth]{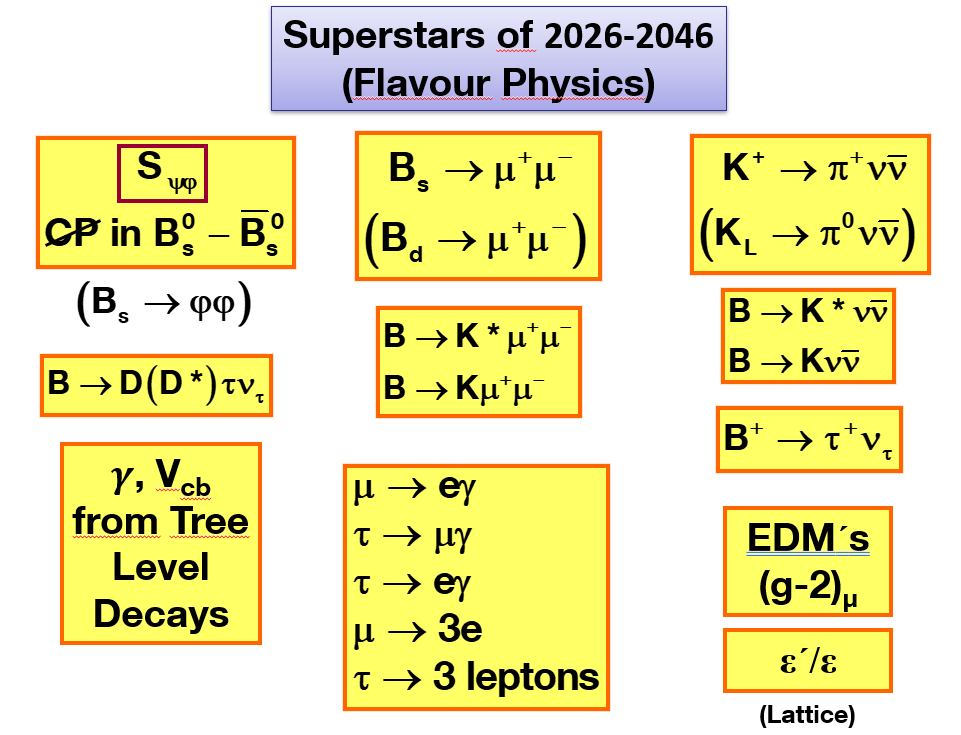}%
\caption{\it Coming Years.\label{Great1}}
\end{figure}

\subsection{Outlook}
Assuming that a least few of these entries will be realized in the coming 20 years, in particular the one on $\epe$, we should have then great time until 2046
 and this is expressed
in Fig.~\ref{Great1}.

Unfortunately, it could take longer time. After all Enrico Fermi, already in the
1930s could sense something going on which was hidden in his $G_F$. But it
took first 30 years to suggest that it could be a heavy gauge boson $W^\pm$
behind these effects and another 20 years to discover it.
Let us hope that it will not take such a long time now and that we
will be able to narrow the number of possible extensions of the SM 
at least through indirect searches until 2046. The methods presented here and
in \cite{Buras:2026vbp}
will hopefully play an important role in these
searches. But from today's  perspective it is difficult
to predict how our field will develop in the coming decades. Indeed, let me
cite Niels Bohr: ``Prediction is very difficult, especially if it's about the future''.

Still let me make the prediction that Andrzej Białas  will be at his 100th
birthday in person. I wish him all the best for the next ten years and beyond.

\vspace{0.4cm}
{\bf Acknowledgements} 

I would like to thank Michal Praszalowicz for inviting me to the Cracow School of 2026. Many thanks go to my collaborators to whom  my Flavour Autobiography \cite{Buras:2026vbp} is dedicated.
Financial support by the Excellence Cluster ORIGINS, funded by the Deutsche Forschungsgemeinschaft (DFG, German Research Foundation), Excellence Strategy, EXC-2094, 390783311 is acknowledged.

        \renewcommand{\refname}{R\lowercase{eferences}}

\addcontentsline{toc}{section}{References}
 
\bibliographystyle{JHEP}
\bibliography{flavour}
\end{document}